\documentclass[a4paper,12pt]{article}
\usepackage{a4,amsmath,amssymb,axodraw,slashed,epsf}
\usepackage[square,comma,numbers,sort&compress]{natbib}
\usepackage[dvips]{color}
\usepackage[dvips]{xcolor}
\usepackage{hyperref}
\usepackage{cancel}
\usepackage{ulem}

\begin{document}


\newcommand{\bed}{\begin{displaymath}}
\newcommand{\eed}{\end{displaymath}}
\newcommand{\beq}{\begin{equation}}
\newcommand{\eeq}{\end{equation}}
\newcommand{\bea}{\begin{eqnarray}}
\newcommand{\eea}{\end{eqnarray}}
\newcommand{\tgb}{{\rm tg}\beta}
\newcommand{\tga}{{\rm tg}\alpha}
\newcommand{\stgb}{{\rm tg}^2\beta}
\newcommand{\sia}{\sin\alpha}
\newcommand{\coa}{\cos\alpha}
\newcommand{\sib}{\sin\beta}
\newcommand{\cob}{\cos\beta}
\newcommand{\MS}{\overline{\rm MS}}
\newcommand{\st}{\tilde{t}}
\newcommand{\sgl}{\tilde{g}}
\newcommand{\Dmb}{\ensuremath{\Delta_b}}

\newcommand{\tb}{{\rm tg}\beta}
\newcommand{\gluino}{\widetilde{g}}
\newcommand{\squark}{\tilde{q}}
\newcommand{\Mq}[1]{m_{\tilde{q}_{#1}}}
\newcommand{\mq}{m_{q}}
\newcommand{\Aq}{A_q}
\newcommand{\mix}{\widetilde{\theta}}
\newcommand{\Lag}{\mathcal{L}}
\newcommand{\ls}{\lambda_{s}}
\newcommand{\lb}{\lambda_{b}}
\newcommand{\lt}{\lambda_t}
\newcommand{\Deltamb}{\Delta_b}
\newcommand{\DeltambQCD}{\Delta_b^{QCD}}
\newcommand{\DeltambELW}{\Delta_b^{elw}}
\newcommand{\mb}{m_{b}}
\newcommand{\mt}{m_{t}}
\newcommand{\bR}{b_R}
\newcommand{\bL}{b_L}
\newcommand{\sR}{s_R}
\newcommand{\sL}{s_L}
\newcommand{\At}{A_t}
\newcommand{\Sb}{\Sigma_b}
\newcommand{\Ao}[1]{A_0(#1)}
\newcommand{\Co}[5]{C_0(#1,#2;#3,#4,#5)}
\newcommand{\as}{\alpha_s}
\newcommand{\Mg}{m_{\tilde{g}}}
\newcommand{\Mb}[1]{m_{\tilde{b}_{#1}}}
\newcommand{\Mt}[1]{m_{\tilde{t}_{#1}}}
\newcommand{\NL}{\nonumber\\}
\newcommand{\bs}{\widetilde{b}}
\newcommand{\str}{\widetilde{s}}
\newcommand{\bsL}{\bs_L}
\newcommand{\bsR}{\bs_R}
\newcommand{\strL}{\str_L}
\newcommand{\strR}{\str_R}
\newcommand{\ts}{\widetilde{t}}
\newcommand{\tsR}{\widetilde{t}_R}
\newcommand{\tsL}{\widetilde{t}_L}
\newcommand{\higgsino}{\widetilde{h}}
\newcommand{\stopx}{\tilde{t}}
\newcommand{\sbottom}{\tilde{b}}
\newcommand{\strange}{\tilde{s}}
\newcommand{\gh}{g_b^h}
\newcommand{\gH}{g_b^H}
\newcommand{\gA}{g_b^A}
\newcommand{\ght}{\tilde{g}_b^h}
\newcommand{\gHt}{\tilde{g}_b^H}
\newcommand{\gAt}{\tilde{g}_b^A}
\newcommand{\stb}{{\rm tg}^2\beta}
\newcommand{\Order}[1]{{\cal{O}}\left(#1\right)}
\newcommand{\Msusy}{M_{SUSY}}
\newcommand{\gluon}{g}
\newcommand{\sttop}{\widetilde{t}}
\newcommand{\CF}{C_F}
\newcommand{\CA}{C_A}
\newcommand{\TR}{T_R}
\newcommand{\Tr}[1]{{\rm Tr}\left[#1\right]}
\newcommand{\dk}{\frac{d^nk}{(2\pi)^n}}
\newcommand{\dqq}{\frac{d^nq}{(2\pi)^n}}
\newcommand{\partderiv}[1]{\frac{\partial}{\partial#1}}
\newcommand{\kmu}{k_{\mu}}
\newcommand{\qmu}{q_{\mu}}
\newcommand{\eUV}{\epsilon}
\newcommand{\dMq}[1]{\delta\Mq{#1}}
\newcommand{\gs}{g_s}
\newcommand{\MSbar}{\overline{\rm MS}}
\newcommand{\gPhit}{\tilde{g}_b^{\Phi}}
\newcommand{\muR}{\mu_{R}}
\newcommand{\bbbar}{b\bar{b}}
\newcommand{\MPhi}{M_{\Phi}}
\newcommand{\dlt}{\delta\lambda_t}
\newcommand{\GF}{{\rm G_F}}
\newcommand{\gPhi}{g_b^{\Phi}}
\newcommand{\mbMS}{\overline{m}_{b}}
\newcommand{\asrun}[1]{\as(#1)}
\newcommand{\NF}{N_F}
\newcommand{\gtPhi}{g_t^{\Phi}}
\newcommand{\Log}[1]{\log\left(#1\right)}
\newcommand{\smallaeff}{small $\alpha_{eff}$}
\newcommand{\ra}{\rightarrow}
\newcommand{\tautau}{\tau^+\tau^-}

\newcommand{\Quark}[4]{\ArrowLine(#1,#2)(#3,#4)}
\newcommand{\Gluino}[5]{\Gluon(#1,#2)(#3,#4){3}{#5}\Line(#1,#2)(#3,#4)}
\newcommand{\CrossedCircle}[6]{\BCirc(#1,#2){5}\Line(#3,#6)(#5,#4)\Line(#5,#6)(#3,#4)}
\newcommand{\Higgsino}[5]{\Photon(#1,#2)(#3,#4){3}{#5}\Line(#1,#2)(#3,#4)}
\newcommand{\myGluon}[5]{\Gluon(#1,#2)(#3,#4){3}{#5}}
\newcommand{\Squark}[4]{\DashArrowLine(#1,#2)(#3,#4){2}}
\newcommand{\Squarknoarrow}[4]{\DashLine(#1,#2)(#3,#4){2}}
\newcommand{\Quarknoarrow}[4]{\ArrowLine(#1,#2)(#3,#4)}

\newcommand{\lsim}{\raisebox{-0.13cm}{~\shortstack{$<$ \\[-0.07cm] $\sim$}}~}
\newcommand{\gsim}{\raisebox{-0.13cm}{~\shortstack{$>$ \\[-0.07cm] $\sim$}}~}

\newcommand{\trex}[1]{\textcolor{red}{#1}}
\newcommand{\SC}[1]{\textcolor{cyan}{#1}}
\newcommand{\SCc}[1]{\textcolor{cyan}{[SC: #1]}}
\newcommand{\JC}[1]{\textcolor{green}{#1}}
\newcommand{\AB}[1]{\textcolor{blue}{#1}}
\newcommand{\JR}[1]{\textcolor{orange}{#1}}
\newcommand{\MM}[1]{\textcolor{violet}{#1}}


\vspace*{-2.5cm}

\begin{flushright}
KA--TP--35--2025
\end{flushright}

\begin{center}
{\large \sc Higgs-Pair Production via Gluon Fusion: Top-Yukawa- \\[0.2cm]
and light-quark-induced electroweak Corrections}
\end{center}

\begin{center} {\sc A.~Bhattacharya$^1$, F.~Campanario$^1$,
S.~Carlotti$^2$, J.~Chang$^{3,4}$, J.~Mazzitelli$^3$,
M.~M\"uhlleitner$^2$, J.~Ronca$^5$, and M.~Spira$^3$} \\[0.8cm]

\begin{small}
{\it \small
$^1$ Theory Division, IFIC, University of Valencia-CSIC, E--46980 Paterna,
  Valencia, Spain \\
$^2$ Institute for Theoretical Physics, KIT, D--76128 Karlsruhe, Germany \\
$^3$ PSI Center for Neutron and Muon Sciences, CH--5232 Villigen PSI,
Switzerland \\
$^4$ Institute for Theoretical Physics, ETH Z\"urich, CH--8093 Z\"urich,
Switzerland \\
$^5$ Dipartimento di Fisica e Astronomia, Universit\`a degli Studi di Padova,
and INFN, Sezione di Padova, Via Marzolo 8, I--35131 Padova, Italy}
\end{small}
\end{center}


\begin{abstract}
\noindent
Gluon fusion, $gg\to HH$, is the dominant Higgs-pair production process
at the Large Hadron Collider (LHC) and provides the first
direct access to the trilinear Higgs self-interaction. The process is
loop-induced, with the main contribution emerging from top-quark
loops within the Standard Model. In the past, the QCD corrections have
been calculated and found to increase the cross section significantly.
With the anticipated accuracies achievable at the high-luminosity LHC
(HL--LHC), the theoretical uncertainties will be of increased relevance to
compete with the experimental precision at the level of less than 30\%. In
this work, we take the next steps towards the determination of the complete
electroweak corrections at next-to-leading order by calculating
the full top-Yukawa and light-quark induced corrections. These
corrections modify the cross section moderately in the kinematical
regimes of interest.
\end{abstract}


\section{Introduction}
After the discovery of the Higgs boson in 2012 \cite{discovery}, increasingly precise measurements have been made to determine its properties \cite{couplings}. 
Early experiments confirmed that its spin and CP were consistent with the spinless and CP-even nature of the Standard Model (SM) Higgs boson. 
Subsequent measurements of the boson’s couplings to the top and bottom quarks, the $\tau$ lepton, and the electroweak gauge bosons, $W$ and $Z$, further supported the SM-like nature by showing agreement with the SM predictions within their respective experimental uncertainties. 
Moreover, the loop-induced couplings to photons and gluons have been determined, and recently, evidence for Higgs couplings to second-generation muons and to $Z\gamma$ has been found \cite{h2zgamma}. 
So far, no deviations from the properties of the SM Higgs boson have been observed. 
In addition to probing the Higgs boson couplings to muons
and charm quarks, one of the major goals of the HL--LHC is the
measurement of the trilinear Higgs self-coupling $\lambda_{HHH}$, as the
first step towards the experimental reconstruction of the scalar Higgs
potential \cite{hhpro}.

The Higgs sector of the SM consists of a scalar isospin doublet
containing three would-be Goldstone bosons and one physical Higgs field
\cite{higgs}. The latter acquires a finite vacuum expectation value
resulting in a non-trivial realization of the electroweak gauge
symmetry. The fluctuations of the scalar Higgs field around this vacuum
expectation value describe the Higgs boson as part of the physical
states of the SM. The Higgs boson allows the gauge symmetry to be
unbroken, so that the electroweak interactions are weak up to high-energy
scales \cite{unitarity} and crucially, renormalizable \cite{smren}. At
the LHC, the SM Higgs boson is dominantly produced in the loop-induced
gluon fusion process $gg\to H$, while all the other production modes,
vector-boson fusion $qq\to qqH$, Higgs-strahlung $q\bar q\to W/Z + H$
and bremsstrahlung off top quarks $gg/q\bar q\to t\bar t H$, are
suppressed by at least one order of magnitude (see, e.g.~\cite{review}).

The trilinear Higgs self-coupling will first become directly accessible
in Higgs-boson pair production \cite{hhrev}. The dominant Higgs-boson
pair production channel is gluon fusion $gg\to HH$, while the other
processes, vector-boson fusion $qq\to qqHH$, double Higgs-strahlung
$q\bar q\to HH + W/Z$ and associated production with a $t\bar t$
pair $gg,q\bar q\to t\bar t H$, are suppressed by an order of magnitude
or more. The pair production processes are suppressed by three orders of magnitude compared to their single Higgs counterparts.  In this work, we focus on
the dominant gluon fusion process. The leading order (LO) cross section
was calculated a long time ago \cite{gghhlo} and the QCD corrections
were calculated at next-to-leading order (NLO) \cite{nlohtl},
next-to-NLO (NNLO) \cite{nnlohtl} and next-to-NNLO (N$^3$LO)
\cite{n3lohtl} in the heavy-top limit (HTL). These contributions
increase the cross section by more than a factor of two.
More recently, the full NLO QCD corrections, including the full top-quark mass dependence, have become available. These corrections reduce the total cross section by approximately 15\% relative to the Born-improved HTL, although their impact is more pronounced in differential distributions \cite{gghhnlo,gghhnlo1,gghhqcd2}.
These calculations were performed by
applying two different methods of numerical integration.  This has been
supported by suitable expansion methods in the low- and high-energy
regimes, as well as covering the full kinematical range
\cite{gghhexp,gghhexp1}. The complete NLO results were combined with the
NNLO corrections in the HTL with the full mass dependence included in
the double-real corrections \cite{nnloftapprox}. In addition, soft-gluon
resummation effects have been studied up to
next-to-next-to-next-to-leading logarithmic order (N$^3$LL) \cite{gghhresum}.
The grids of the full numerical calculation were implemented in {\tt POWHEG-BOX} and {\tt aMC@NLO} \cite{gghhpowheg} and, recently, in the Geneva
generator \cite{gghhgeneva}. In addition, the expansion approach
\cite{gghhexp1} was linked to {\tt POWHEG-BOX} so that it yields reliable
results at the per-cent-level accuracy in the full kinematical range.
However, the theoretical uncertainties were found to be dominated by
the scale and scheme dependence of the virtual top mass and amount to a maximum of 25\% for the total cross section. These uncertainties are larger for
kinematical distributions \cite{gghhnlo1,gghhqcd2,gghhexp1}.  Dedicated
efforts are required to reduce them to at-most the level of 10\%.  The
full NNLO QCD corrections are partly known meanwhile \cite{gghhnnlo},
but are not yet complete. Another approach has analyzed
the leading logarithmic structure of the cross section by using SCET
methods at high energies \cite{gghhscet}. A reduction of the scale and
scheme dependence at high energy, however, requires the analysis to be
extended to the subleading logarithmic level and to the low-energy regime
close to the production threshold and the virtual $t\bar t$-threshold,
which is beyond the work of Ref.~\cite{gghhscet}.

The next step in reducing the theoretical uncertainties is to calculate the
NLO electroweak corrections. The first work in this direction has
determined the top-Yukawa-induced corrections in the HTL for the
effective Higgs couplings to gluons, while the trilinear Higgs-vertex
corrections have been included with full mass dependence \cite{schlenk}.
This analysis has pointed out that the use of a radiatively
corrected effective trilinear Higgs coupling is disfavored upon comparison
with the results using the LO-like coupling. 
The first complete electroweak corrections to Higgs boson pair production have been obtained in Ref.~\cite{gghhelw}. These corrections reach the $-10\%$ level at large invariant Higgs-pair masses $M_{HH}$, while more pronounced effects appear near the production threshold $M_{HH}\gsim 2M_H$ due to the large destructive interference effects of the LO matrix element, rendering this regime as sub-leading for the physical observable. Although the results of Ref.\ \cite{gghhelw} are technically comprehensive, a decomposition into individually identifiable contributions is not yet known. Consequently, further analytic insight into the underlying mechanisms remains valuable for understanding the origin and kinematic features of the electroweak corrections. Recently, a
 complete numerical calculation of the top-Yukawa- and Higgs
self-interaction-induced corrections has been presented, where only
Higgs-exchange diagrams have been taken into account \cite{gghhhiggs}.
For large values of $M_{HH}$, analytical high-energy expansions exist,
which confirm the small size of the top-Yukawa-induced corrections
\cite{gghhelwexp}. The contribution of light-quark loops has also been derived and found to be small \cite{bonetti}.

This work presents the first dedicated numerical evaluation of the full top-Yukawa-induced corrections involving Higgs and Goldstone boson exchange diagrams, as well as the light-quark loop contributions. These results provide a physically transparent decomposition of the electroweak effects, complementing existing calculations by isolating and quantifying the contributions of individual diagram classes.
In Section \ref{sc:lo}, we review the LO results
and define our notation. Section \ref{sc:elw} describes our method to
perform the numerical integration and renormalization, while in Section
\ref{sc:results} we present our results. In Section
\ref{sc:conclusions}, we conclude this work by summarizing our findings.

\section{Higgs-boson pair production via gluon fusion} 
\label{sc:lo}
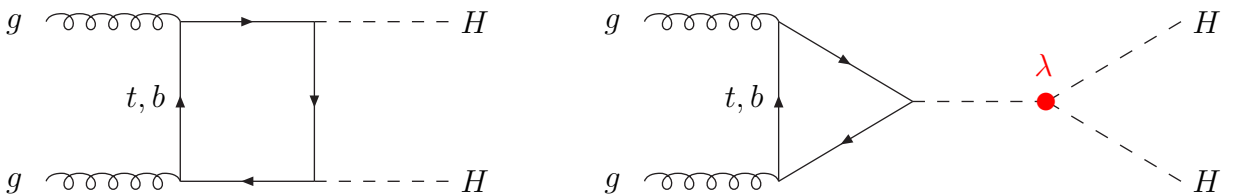
\begin{figure}[hbtp]
\begin{center}
\setlength{\unitlength}{1pt}
\begin{picture}(100,90)(100,0)
\Gluon(0,80)(50,80){3}{5}
\Gluon(0,20)(50,20){3}{5}
\ArrowLine(50,20)(50,80)
\ArrowLine(50,80)(100,80)
\ArrowLine(100,80)(100,20)
\ArrowLine(100,20)(50,20)
\DashLine(100,80)(150,80){5}
\DashLine(100,20)(150,20){5}
\put(155,76){$H$}
\put(155,16){$H$}
\put(-15,78){$g$}
\put(-15,18){$g$}
\put(30,48){$t,b$}
\end{picture}
\begin{picture}(100,90)(-20,0)
\Gluon(0,80)(50,80){3}{5}
\Gluon(0,20)(50,20){3}{5}
\ArrowLine(50,20)(50,80)
\ArrowLine(50,80)(100,50)
\ArrowLine(100,50)(50,20)
\DashLine(100,50)(150,50){5}
\DashLine(150,50)(200,80){5}
\DashLine(150,50)(200,20){5}
\put(205,76){$H$}
\put(205,16){$H$}
\put(-15,78){$g$}
\put(-15,18){$g$}
\put(30,48){$t,b$}
\put(146,46){\Large \textcolor{red}{$\bullet$}}
\put(146,60){\textcolor{red}{$\lambda$}}
\end{picture}
\setlength{\unitlength}{1pt}
\caption{\label{fg:hhdialo} \it Typical diagrams of Higgs-boson
pair production via gluon fusion at LO. The contribution of the trilinear
Higgs coupling is marked in red.}
\end{center}
\end{figure}
\noindent
At LO, Higgs pair production is generated primarily by top-quark loops, with
minor contributions from bottom-quark loops. The contributing diagrams are
shown in Fig.~\ref{fg:hhdialo} and are grouped into two classes, box and triangle diagrams, where the latter contains an $s$-channel Higgs propagator
and the trilinear Higgs coupling $\lambda = \lambda_{HHH}$. At this order, the
total and differential production cross sections with respect to the
invariant Higgs-pair mass are given by
\begin{eqnarray}
\sigma_{\mathrm{LO}} & = & \int_{\tau_0}^1 d\tau~\frac{d{\cal
L}^{gg}}{d\tau}~\hat\sigma_{\mathrm{LO}}(Q^2 = \tau s) \nonumber \\[0.5cm]
\frac{d\sigma_{LO}}{dQ^2} & = & \left. \frac{d{\cal
L}^{gg}}{d\tau}~\frac{\hat\sigma_{\mathrm{LO}}(Q^2)}{s} \right|_{\tau =
\frac{Q^2}{s}} \, ,
\end{eqnarray}
where ${\cal L}^{gg}$ denotes the gluonic parton luminosity involving
the gluon densities $g(x,\mu_F)$,
\begin{equation}
\frac{d{\cal L}^{gg}}{d\tau} = \int_\tau^1 \frac{dx}{x} g(x,\mu_F)
g\left(\frac{\tau}{x},\mu_F\right)
\label{eq:lgg}
\end{equation}
at the factorization scale $\mu_F$, and the integration boundary is given
by $\tau_0=4M_H^2/s$, where $s$ denotes the hadronic center-of-mass
(c.m.) energy squared and $M_H$ the Higgs mass. The scale $Q^2=M_{HH}^2$
is defined in terms of the invariant mass $M_{HH}$ of the Higgs pair.
The partonic cross section at LO can be cast into the form
\begin{equation}
\hat\sigma_{LO} = \frac{G_F^2\alpha_s^2(\mu_R)}{512 (2\pi)^3}
\int_{\hat t_-}^{\hat t_+} d\hat t \Big[ | C_\triangle F_\triangle +
F_\Box|^2 + |G_\Box|^2 \Big] \,,
\label{eq:gghhlo}
\end{equation}
where the integration boundaries are given by
\begin{equation}
\hat t_\pm = -\frac{1}{2} \left[ Q^2 - 2M_H^2 \mp Q^2
\sqrt{1-4\frac{M_H^2}{Q^2}} \right] \, ,
\label{eq:tbound}
\end{equation}
and the symmetry factor 1/2 for the identical Higgs bosons in the
final state is included. The coefficient $C_\triangle = v\lambda_{HHH}
/(Q^2-M_H^2)$ contains the LO trilinear Higgs coupling
\begin{equation}
\lambda_{HHH} = 3\frac{M_H^2}{v} \,,
\label{eq:lambdaLO}
\end{equation}
that is related to the Higgs mass and the vacuum expectation value (vev) $v$,
where the vev is connected to the Fermi constant $G_F = 1/(\sqrt{2}
v^2)$. The factor $\alpha_s(\mu_R)$ denotes the strong coupling at the
renormalization scale $\mu_R$.  The form factors $F_\triangle$ of the LO
triangle diagrams and $F_\Box, G_\Box$ of the LO box diagrams can be
found in Refs.~\cite{gghhlo}. In the HTL, they approach simple
expressions: $F_\triangle \to 2/3$, $F_\Box \to -2/3$ and $G_\Box\to 0$.

\section{Electroweak corrections} \label{sc:elw}
In this section, we discuss the top-Yukawa and light-quark induced
electroweak corrections, with the latter neglecting the virtual quark
masses. Since the electroweak corrections do not  involve real
corrections, their total sum is infrared and collinear finite. Therefore, the NLO corrections can be accommodated by a shift of the corresponding
LO form factors
\begin{equation}
F_\triangle \to F_\triangle~(1+\Delta_\triangle), \quad F_\Box \to
F_\Box~ (1+\Delta_\Box), \quad G_\Box\to G_\Box~(1+\Delta_G) \, .
\label{eq:corr}
\end{equation}
From this, one can expand Eq.~(\ref{eq:gghhlo}) in the electroweak
couplings. At NLO, the result reads
\begin{equation}
\frac{d\sigma_{NLO}}{dQ^2} = \frac{d\sigma_{LO}}{dQ^2}~(1+\delta)
\label{eq:delta}
\end{equation}
with
\begin{equation}
\delta = 2\Re e~\frac{\int_{\hat t_-}^{\hat t_+} d\hat t \Big[
(C_\triangle F_\triangle + F_\Box)^* (C_\triangle F_\triangle
\Delta_\triangle + F_\Box \Delta_\Box) + |G_\Box|^2 \Delta_G \Big]}
{\int_{\hat t_-}^{\hat t_+} d\hat t \Big[ | C_\triangle F_\triangle +
F_\Box|^2 + |G_\Box|^2 \Big]} \,,
\end{equation}
There are two tensor structures that contribute to the matrix element,
corresponding to the total angular-momentum states with $S_z=0$ and $2$,
\begin{eqnarray}
{\cal M}[g^a(q_1) g^b(q_2) \to H(p_1) H(p_2)] & = & -i\,\delta_{ab}\,\frac{G_F\alpha_s(\mu_R)
Q^2}{2\sqrt{2}\pi} {\cal A}^{\mu\nu} \epsilon_{1\mu} \epsilon_{2\nu} \, ,
\nonumber \\[0.3cm]
\mbox{where} \qquad\quad {\cal A}^{\mu\nu} & = & F_1 T_1^{\mu\nu} + F_2
T_2^{\mu\nu} \, , \nonumber \\[0.3cm]
F_1 & = & C_\triangle F_\triangle + F_\Box \, , \qquad
\qquad \qquad
F_2 = G_\Box \, , \nonumber \\[0.3cm]
\mbox{and} \qquad \quad
T_1^{\mu\nu} & = & g^{\mu\nu}-\frac{q_1^\nu q_2^\mu}{(q_1q_2)}\, , \nonumber \\
T_2^{\mu\nu} & = & g^{\mu\nu}+\frac{M_H^2 q_1^\nu q_2^\mu}{p_T^2 (q_1q_2)}
-2\frac{(q_2 p_1) q_1^\nu p_1^\mu}{p_T^2 (q_1q_2)}
-2\frac{(q_1 p_1) p_1^\nu q_2^\mu}{p_T^2 (q_1q_2)}
+2\frac{p_1^\nu p_1^\mu}{p_T^2} \, , \nonumber \\
\mbox{with} \qquad p_T^2 & = & 2 \frac{(q_1 p_1)(q_2 p_1)}{(q_1 q_2)} -
M_H^2 \, ,
\label{eq:ff}
\end{eqnarray}
where $p_T$ is the transverse momentum of each of the final-state Higgs
bosons. Working in $n=4-2\epsilon$ dimensions, the following projectors
can be constructed on the two form factors,
\begin{equation}
P_1^{\mu\nu} = \frac{(1-\epsilon) T_1^{\mu\nu} + \epsilon
T_2^{\mu\nu}}{2(1-2\epsilon)}\, , \qquad \qquad
P_2^{\mu\nu} = \frac{\epsilon T_1^{\mu\nu} + (1-\epsilon)
T_2^{\mu\nu}}{2(1-2\epsilon)} \, ,
\label{eq:projector}
\end{equation}
such that
\begin{equation}
P_1^{\mu\nu} {\cal A}_{\mu\nu} = F_1 \, , \qquad \qquad
P_2^{\mu\nu} {\cal A}_{\mu\nu} = F_2 \, .
\end{equation}
Using these projectors, the explicit results of the two form factors
$F_{1,2}$ can be obtained in a straightforward manner at LO and NLO.

\subsection{Top--Yukawa-induced electroweak corrections}
In order to define the top-Yukawa induced electroweak corrections
consistently, we work in the gaugeless limit for these contributions
\cite{gaugeless}.  This is defined by switching off electroweak gauge
interactions and working in a QCD-improved top+bottom Yukawa model,
with the Lagrangian
\begin{eqnarray}
{\cal L} & = & -\frac{1}{4} G^{a\mu\nu} G^a_{\mu\nu} + \bar t i \!\!
\not\!\! D t  + \bar b i \!\! \not\!\! D b + |\partial_\mu \phi|^2 - V(\phi) - g_t \bar Q_L \phi^c t_R
- g_b \bar Q_L \phi b_R \\
V(\phi) & = & -\mu^2 |\phi|^2 + \frac{\lambda}{2} |\phi|^4 \nonumber \\
& = & -\frac{M_H^2}{8} v^2 + \frac{M_H^2}{2}H^2 + \frac{M_H^2}{v} \left[
\frac{H^3}{2} + \frac{H}{2} (G^0)^2 + HG^+G^-\right] \nonumber \\
& + & \!\!\!\!\! \frac{M_H^2}{2v^2}
\left[ \frac{H^4}{4} + \frac{H^2}{2} (G^0)^2 + H^2 G^+ G^- + (G^+G^-)^2 +
(G^0)^2 G^+G^- + \frac{(G^0)^4}{4} \right]\, .
\end{eqnarray}
Here, we define the Yukawa couplings as $g_{t/b} = \sqrt{2} m_{t/b}/v$, and the covariant derivative $D_\mu$ with respect to QCD. The left-handed quark isospin doublet is given by
\begin{eqnarray}
Q_L = \left( \begin{array}{c} t \\ \displaystyle b \end{array}
\right)_L \, ,
\end{eqnarray}
with $t_R$ denoting the right-handed top-quark field. The scalar Higgs isodoublet is defined as\footnote{$\phi^c = i\sigma^2 \phi^*$ denotes the charge-conjugated Higgs doublet.}
\begin{eqnarray}
\phi = \left( \begin{array}{c} 
                G^+ \\ \displaystyle 
                \frac{v+H+iG^0}{\sqrt{2}} 
              \end{array} \right),
\end{eqnarray}
where $H$ is the physical Higgs field and $G^\pm$, $G^0$ are the charged and neutral Goldstone bosons, respectively. The Higgs potential is denoted by $V(\phi)$. The vacuum expectation value
is given by $v^2 = 2\mu^2/\lambda$ and the Higgs mass by $M_H^2 =
\lambda v^2$ in terms of the two initial parameters $\mu^2, \lambda$ of
the Higgs potential.  Within the gaugeless limit, the Goldstones are
massless,
\begin{eqnarray}
M_{G^0} = M_{G^\pm} = 0 \, .
\end{eqnarray}
The neutral and charged Goldstones couple to the top quark with the top
Yukawa coupling, too, and thus need to be taken into account to define the
full top-Yukawa induced corrections consistently\footnote{In unitary
gauge of the full electroweak sector, the top-Yukawa coupling of the
would-be Goldstones is hidden in the contributions of the longitudinal
$W,Z$ components and needs to be extracted by suitable expansions in the
large top-mass limit for a complete top-Yukawa induced contribution.}.

\subsubsection{Two-loop triangle diagrams}
\begin{figure}[hbt]
\begin{center}
\setlength{\unitlength}{1pt}
\begin{picture}(100,85)(120,0)
\Gluon(0,80)(50,80){3}{5}
\Gluon(0,20)(50,20){3}{5}
\ArrowLine(50,20)(50,80)
\ArrowLine(50,80)(75,65)
\ArrowLine(75,65)(100,50)
\ArrowLine(100,50)(75,35)
\ArrowLine(75,35)(50,20)
\DashLine(75,35)(75,65){5}
\DashLine(100,50)(130,50){5}
\DashLine(130,50)(180,80){5}
\DashLine(130,50)(180,20){5}
\put(185,76){$H$}
\put(185,16){$H$}
\put(55,60){$H,$}
\put(55,47){$G^0,$}
\put(55,34){$G^\pm$}
\put(-15,78){$g$}
\put(-15,18){$g$}
\put(30,48){$t/b$}
\put(85,65){$t$}
\end{picture}
\begin{picture}(100,85)(-20,0)
\Gluon(0,80)(50,80){3}{5}
\Gluon(0,20)(50,20){3}{5}
\ArrowLine(50,20)(50,50)
\ArrowLine(50,50)(50,80)
\ArrowLine(50,80)(75,65)
\ArrowLine(75,65)(100,50)
\ArrowLine(100,50)(50,20)
\DashLine(50,50)(75,65){5}
\DashLine(100,50)(130,50){5}
\DashLine(130,50)(180,80){5}
\DashLine(130,50)(180,20){5}
\put(185,76){$H$}
\put(185,16){$H$}
\put(55,45){$H,G^0,$}
\put(55,35){$G^\pm$}
\put(-15,78){$g$}
\put(-15,18){$g$}
\put(30,65){$t/b$}
\put(40,35){$t$}
\end{picture} \\
\begin{picture}(100,90)(120,0)
\Gluon(0,80)(50,80){3}{5}
\Gluon(0,20)(50,20){3}{5}
\ArrowLine(50,20)(50,50)
\ArrowLine(50,50)(50,80)
\ArrowLine(50,80)(100,50)
\ArrowLine(100,50)(75,35)
\ArrowLine(75,35)(50,20)
\DashLine(50,50)(75,35){5}
\DashLine(100,50)(130,50){5}
\DashLine(130,50)(180,80){5}
\DashLine(130,50)(180,20){5}
\put(185,76){$H$}
\put(185,16){$H$}
\put(55,60){$H,$}
\put(55,47){$G^0,G^\pm$}
\put(-15,78){$g$}
\put(-15,18){$g$}
\put(40,65){$t$}
\put(30,35){$t/b$}
\end{picture}
\begin{picture}(100,90)(-20,0)
\Gluon(0,80)(50,80){3}{5}
\Gluon(0,20)(50,20){3}{5}
\ArrowLine(50,20)(50,80)
\ArrowLine(50,80)(100,50)
\ArrowLine(100,50)(50,20)
\DashCArc(75,65)(15,-32,148){5}
\DashLine(100,50)(130,50){5}
\DashLine(130,50)(180,80){5}
\DashLine(130,50)(180,20){5}
\put(185,76){$H$}
\put(185,16){$H$}
\put(75,85){$H,G^0,G^\pm$}
\put(-15,78){$g$}
\put(-15,18){$g$}
\put(40,45){$t$}
\put(65,50){$t/b$}
\end{picture} \\
\begin{picture}(100,90)(120,0)
\Gluon(0,80)(50,80){3}{5}
\Gluon(0,20)(50,20){3}{5}
\ArrowLine(50,20)(50,80)
\ArrowLine(50,80)(100,50)
\ArrowLine(100,50)(50,20)
\DashCArc(75,35)(15,-150,30){5}
\DashLine(100,50)(130,50){5}
\DashLine(130,50)(180,80){5}
\DashLine(130,50)(180,20){5}
\put(185,76){$H$}
\put(185,16){$H$}
\put(75,7){$H,G^0,G^\pm$}
\put(-15,78){$g$}
\put(-15,18){$g$}
\put(40,45){$t$}
\put(65,40){$t/b$}
\end{picture}
\begin{picture}(100,90)(-20,0)
\Gluon(0,80)(50,80){3}{5}
\Gluon(0,20)(50,20){3}{5}
\ArrowLine(50,20)(50,80)
\ArrowLine(50,80)(100,50)
\ArrowLine(100,50)(50,20)
\DashCArc(50,50)(15,90,270){5}
\DashLine(100,50)(130,50){5}
\DashLine(130,50)(180,80){5}
\DashLine(130,50)(180,20){5}
\put(185,76){$H$}
\put(185,16){$H$}
\put(-20,45){$H,G^0,G^\pm$}
\put(-15,78){$g$}
\put(-15,18){$g$}
\put(65,75){$t$}
\put(55,45){$t/b$}
\end{picture} \\[-0.3cm]
\setlength{\unitlength}{1pt}
\caption{\label{fg:triangle_t} \it Two-loop triangle diagrams of the
top-Yukawa-induced electroweak corrections to Higgs-boson pair
production involving Higgs $H$ and Goldstone $G^0,G^\pm$ exchanges. The
bottom propagators only contribute to the diagrams with charged
Goldstone exchange.}
\end{center}
\end{figure}
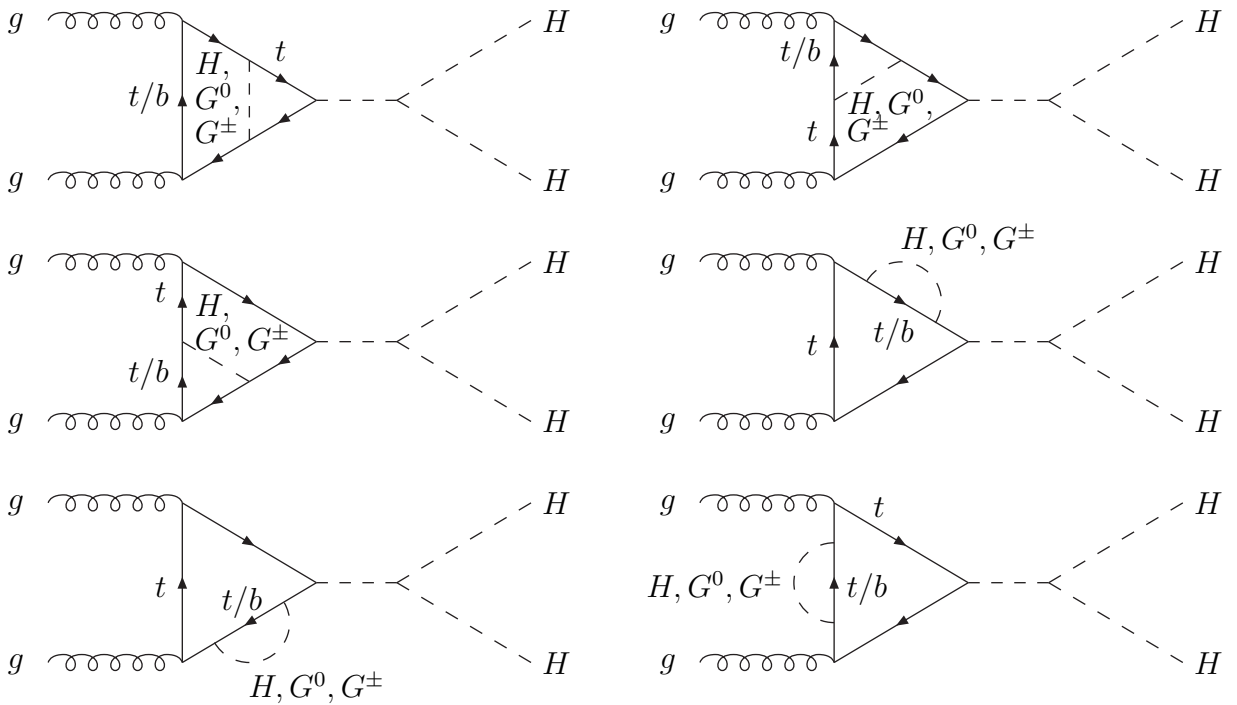

The two-loop diagrams contributing to the top-Yukawa induced component of
$\Delta_\triangle$ can be split into one-loop times one-loop diagrams, and triangle and box diagrams which generate the single- and double-Higgs
couplings to gluons. The two-loop box diagrams will be discussed in the
next paragraph. The one-loop times one-loop diagrams have been
discussed in Ref.~\cite{schlenk} and lead to the correction
$\Delta_{HHH}$, while the two-loop triangle diagrams, depicted in Fig.~\ref {fg:triangle_t}, result in the correction term
$\delta_1$,
\begin{equation}
\Delta_\triangle = \Delta_{HHH} + \delta_1 \, .
\label{eq:triacorr}
\end{equation}
The analytical result of $\Delta_{HHH}$ can be found in
Ref.~\cite{schlenk}.

For the two-loop triangle diagrams we have used the same method as
described in Ref.~\cite{gghhqcd2}, i.e.~we introduced suitable Feynman
parametrizations of each diagram individually, performed end-point
subtractions to separate the divergences, and applied integrations by
parts of the integrands to reduce the power of the denominators, which stabilize the numerical integrations above the individual virtual thresholds. In all diagrams, we encounter virtual $t\bar t$ and $t\bar tH$ thresholds. For charged Goldstone
exchanges, there is an additional virtual $b\bar b$ threshold which
induces the largest instabilities of the three. To
regularize the virtual thresholds we have introduced small imaginary
parts of the propagator masses,
\begin{eqnarray}
m_{t/b}^2 \to m_{t/b}^2 (1-i\bar\epsilon), \quad M_H^2 \to M_H^2
(1-i\bar\epsilon)
\label{eq:imaginary}
\end{eqnarray}
and performed Richardson extrapolations to obtain the limit
$\bar\epsilon \to 0$.

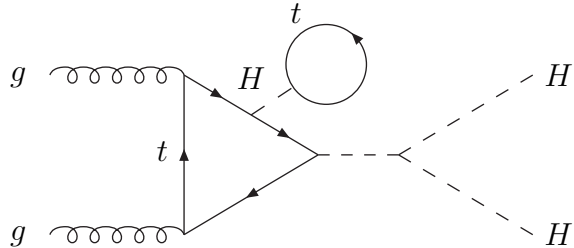
\begin{figure}[hbtp]
\begin{center}
\setlength{\unitlength}{1pt}
\begin{picture}(100,100)(50,0)
\Gluon(0,80)(50,80){3}{5}
\Gluon(0,20)(50,20){3}{5}
\ArrowLine(50,20)(50,80)
\ArrowLine(50,80)(75,65)
\ArrowLine(75,65)(100,50)
\ArrowLine(100,50)(50,20)
\DashLine(75,65)(90,75){5}
\ArrowArc(103,84)(15,-137,223)
\DashLine(100,50)(130,50){5}
\DashLine(130,50)(180,80){5}
\DashLine(130,50)(180,20){5}
\put(185,76){$H$}
\put(185,16){$H$}
\put(70,75){$H$}
\put(-15,78){$g$}
\put(-15,18){$g$}
\put(40,48){$t$}
\put(90,100){$t$}
\end{picture} \\[-0.3cm]
\setlength{\unitlength}{1pt}
\caption{\label{fg:tadpole} \it Example of a two-loop tadpole diagram
of the top-Yukawa induced electroweak corrections to Higgs-boson pair
production involving Higgs exchange. Goldstone exchange does not
contribute to the tadpoles.}
\end{center}
\end{figure}

The counterterm $\delta_{1,CT}$ for the two-loop triangle diagrams
consists of the counterterms for the on-shell Higgs wave
function\footnote{Half of this counterterm is already contained in the
correction $\Delta_{HHH}$.}, the vacuum expectation
value\footnote{This counterterm is determined from the muon decay in the
$G_F$ scheme.} and the top mass that is renormalized on-shell ($x_t
= G_F m_t^2/(8\sqrt{2} \pi^2)$),
\begin{eqnarray}
\delta_{1,CT} & = & \frac{\delta Z_H}{2} - \frac{\delta v}{v} -
2 \delta m_t \frac{1}{F_\triangle} \frac{\partial
F_\triangle}{\partial m_t} \nonumber \\
\delta Z_H = \Sigma'_H(Q^2) & = & 6\, x_t \left\{ (4m_t^2-Q^2) B'_0(Q^2;
m_t, m_t) - B_0(Q^2; m_t, m_t) \right\} \nonumber \\
\frac{\delta v}{v} = \frac{1}{2} \frac{\Sigma_W(0)}{M_W^2} & = &
\frac{T_1}{v M_H^2} + x_t \left\{ B_0(0;m_t,m_b) + 2 B_0(0;m_t,m_t) +
m_t^2 B_0'(0;m_t,m_b) \right\} \nonumber \\
\frac{\delta m_t}{m_t} = \frac{\Sigma(m_t)}{m_t} & = & \frac{T_1}{v
M_H^2} - x_t \left\{ \left.  \left(2-\frac{M_H^2}{2m_t^2} \right)
B_0(p;m_t,M_H) + \frac{m_t^2+m_b^2}{2m_t^2} B_0(p;m_b,0) \right. \right.
\nonumber \\
& & \left. \left. \hspace*{2cm} + \frac{A_0(M_H)-2 A_0(m_t) - A_0(m_b)}{2m_t^2}
\right|_{p^2=m_t^2} \right\}
\end{eqnarray}
where we keep the bottom mass $m_b$ in the arguments of the scalar integrals and the tadpole contribution is given by
\begin{equation}
\frac{T_1}{v} = -12\, x_t A_0(m_t) \, .
\end{equation}
The functions $\Sigma_{W,t}$ are the self-energies of the $W$
boson and the top quark, respectively, while $\Sigma'_H$ denotes the derivative of the
Higgs self-energy with respect to $Q^2$. We have checked explicitly that
the tadpole contributions cancel against the two-loop tadpole diagrams by
introducing tadpole insertions for all of the virtual LO top-quark propagators 
(see Fig.~\ref{fg:tadpole}).  
The scalar integrals $A_0, B_0, C_0$ are the
usual 't Hooft--Veltman one-loop integrals \cite{hooftvelt} and
\begin{equation}
B_0'(q^2; m_1, m_2) = \frac{\partial}{\partial q^2}~B_0(q^2; m_1, m_2)
\, .
\end{equation}
The correction factor $\delta_1$ of Eq.~(\ref{eq:triacorr}) summarizes
the combined result of the two-loop triangle diagrams and counterterms.
Until now, $\delta_1$ has only been calculated in the HTL ($\delta_1 \to
x_t/2$). As a cross-check, we artificially increased the top mass to 3
TeV in our computations and confirmed that $\delta_1$ coincides with
this limit within numerical errors.

\begin{figure}[hbt]
\begin{center}
\begin{picture}(150,245)(0,0)
\put(-150,-155.0){\includegraphics{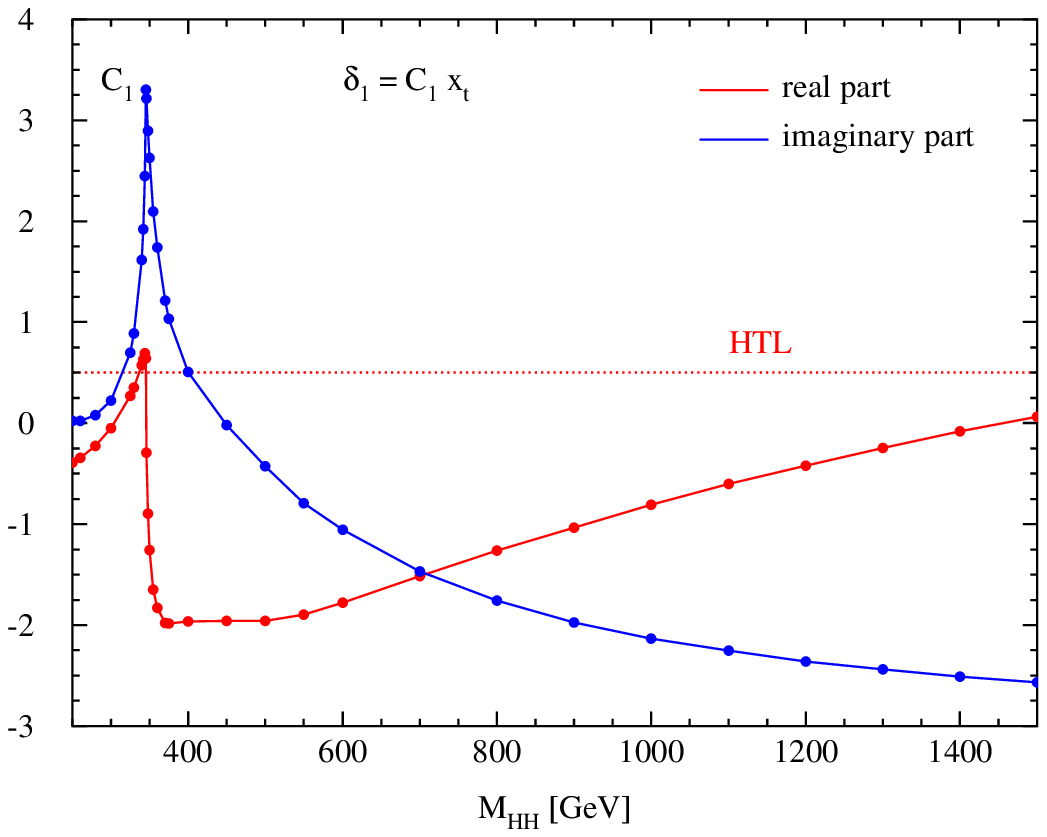}}
\end{picture}
\caption{\it The relative top-Yukawa-induced electroweak correction
factor $\delta_1$ as a function of the invariant Higgs-pair mass
$M_{HH}$ with the top-Yukawa factor $x_t$ stripped off. The dots along
the curves represent the numerical numbers we generated. The dotted
curve shows the HTL of the full result, $C_1\to 1/2$ for large
values of $m_t$.}
\label{fg:d1tria}
\end{center}
\end{figure}

The final result of the correction factor $\delta_1 \equiv C_1 x_t$ is
shown in Fig.~\ref{fg:d1tria} as a function of the invariant Higgs-pair
mass $M_{HH}$. Note that $Q^2 = M_{HH}^2$ is the squared momentum that
flows through the $s$-channel Higgs propagator. Multiplying the shown
factor $C_1$ with $x_t \sim 0.3\%$, the total corrections stay below the
per-cent level.

\subsubsection{Two-loop box diagrams}
\begin{figure}[hbtp]
\begin{center}
\setlength{\unitlength}{1pt}
\begin{picture}(100,90)(100,0)
\Gluon(0,80)(50,80){3}{5}
\Gluon(0,20)(50,20){3}{5}
\ArrowLine(50,20)(50,80)
\ArrowLine(50,80)(100,80)
\ArrowLine(100,80)(100,20)
\ArrowLine(100,20)(50,20)
\DashCArc(50,50)(15,90,270){5}
\DashLine(100,80)(150,80){5}
\DashLine(100,20)(150,20){5}
\put(155,76){$H$}
\put(155,16){$H$}
\put(-20,45){$H,G^0,G^\pm$}
\put(-15,78){$g$}
\put(-15,18){$g$}
\put(105,45){$t$}
\put(55,45){$t/b$}
\end{picture}
\begin{picture}(100,90)(-40,0)
\Gluon(0,80)(50,80){3}{5}
\Gluon(0,20)(50,20){3}{5}
\ArrowLine(50,20)(50,50)
\ArrowLine(50,50)(50,80)
\ArrowLine(50,80)(75,80)
\ArrowLine(75,80)(100,80)
\ArrowLine(100,80)(100,20)
\ArrowLine(100,20)(50,20)
\DashCArc(50,80)(25,-90,0){5}
\DashLine(100,80)(150,80){5}
\DashLine(100,20)(150,20){5}
\put(155,76){$H$}
\put(155,16){$H$}
\put(75,60){$H,$}
\put(55,45){$G^0, G^\pm$}
\put(-15,78){$g$}
\put(-15,18){$g$}
\put(105,45){$t$}
\put(55,85){$t/b$}
\end{picture} \\
\begin{picture}(100,90)(100,0)
\Gluon(0,80)(50,80){3}{5}
\Gluon(0,20)(50,20){3}{5}
\ArrowLine(50,20)(50,50)
\ArrowLine(50,50)(50,80)
\ArrowLine(50,80)(100,80)
\ArrowLine(100,80)(100,50)
\ArrowLine(100,50)(100,20)
\ArrowLine(100,20)(50,20)
\DashLine(50,50)(100,50){5}
\DashLine(100,80)(150,80){5}
\DashLine(100,20)(150,20){5}
\put(155,76){$H$}
\put(155,16){$H$}
\put(60,55){$H,G^0$}
\put(-15,78){$g$}
\put(-15,18){$g$}
\put(105,55){$t$}
\end{picture}
\begin{picture}(100,90)(-40,0)
\Gluon(0,80)(50,80){3}{5}
\Gluon(0,20)(50,20){3}{5}
\ArrowLine(50,20)(50,80)
\ArrowLine(50,80)(75,80)
\ArrowLine(75,80)(100,80)
\ArrowLine(100,80)(100,20)
\ArrowLine(100,20)(75,20)
\ArrowLine(75,20)(50,20)
\DashLine(75,20)(75,80){5}
\DashLine(100,80)(150,80){5}
\DashLine(100,20)(150,20){5}
\put(155,76){$H$}
\put(155,16){$H$}
\put(80,65){$H,$}
\put(80,50){$G^0,$}
\put(80,35){$G^\pm$}
\put(-15,78){$g$}
\put(-15,18){$g$}
\put(105,45){$t$}
\put(30,45){$t/b$}
\end{picture} \\[-0.3cm]
\setlength{\unitlength}{1pt}
\caption{\label{fg:box_t} \it Sample two-loop box diagrams of the
top-Yukawa-induced electroweak corrections to Higgs-boson pair
production involving Higgs $H$ and Goldstone $G^0,G^\pm$ exchanges. The
bottom propagators only contribute to the diagrams with charged
Goldstone exchange.}
\end{center}
\end{figure}
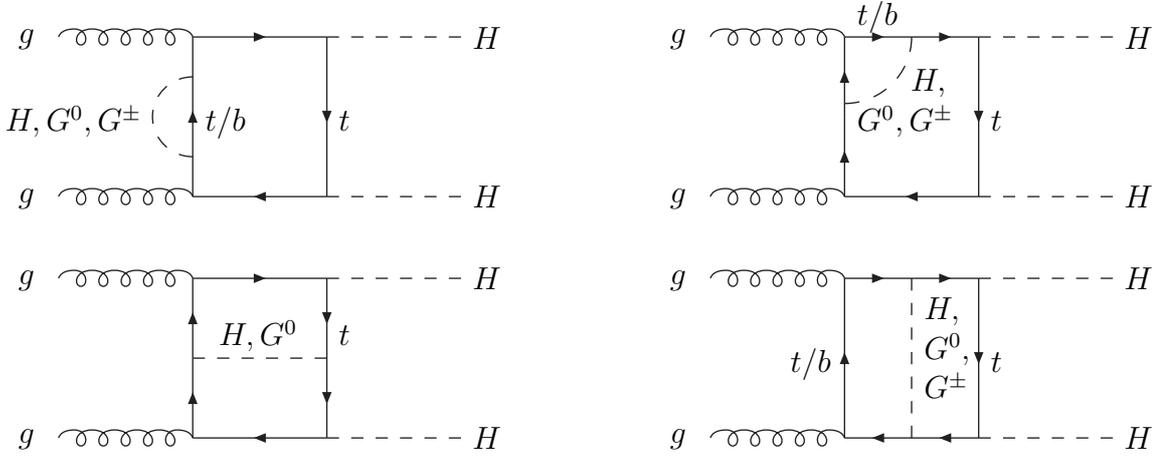

Sample two-loop diagrams describing the genuine box contributions to the
top-Yukawa induced electroweak corrections ($\Delta_\Box, \Delta_G$) are
shown in Fig.~\ref{fg:box_t}.
To obtain the numerical integrals, we apply the projectors of
Eq.~(\ref{eq:projector}) on the two form factors, Feynman
parametrization and end-point subtractions on every diagram.
To stabilize the numerical diagrams above the virtual
thresholds, we performed integration-by-parts and the analytical
continuation of the propagator masses according to Eq.~(\ref{eq:imaginary}).
Finally, Richardson extrapolations are used to reach the
limit $\bar\epsilon \to 0$. In this context, we have modified the
extrapolation polynomials used in Refs.~\cite{gghhnlo1,gghhqcd2} in the
vicinity of the $t\bar t$ threshold, since it was observed in
\cite{ggasqcd} that the $\bar\epsilon$ dependence close to the threshold is not
polynomial in $\bar\epsilon$ but in $\sqrt{\bar\epsilon}$. This
behavior can be
accommodated by modifying the original polynomials displayed in Eq.~(3.33) of
Ref.~\cite{gghhqcd2}\footnote{With the polynomial dependence
$I(\bar\epsilon) = \sum_i c_i \bar\epsilon^{i/2}$ the Richardson
polynomials behave as $R_i(\bar\epsilon) = I(0) + {\cal
O}(\bar\epsilon^{(i+1)/2})$.} to
\begin{eqnarray}
R_1 (\bar\epsilon) & = & \frac{\sqrt{2} I(\bar\epsilon) -
I(2\bar\epsilon)}{\sqrt{2}-1} \nonumber \\
R_2 (\bar\epsilon) & = & \frac{2\sqrt{2} I(\bar\epsilon) - (2+\sqrt{2})
I(2\bar\epsilon) + I(4\bar\epsilon)}{\sqrt{2}-1} \nonumber \\
R_3 (\bar\epsilon) & = & \frac{-8 I(\bar\epsilon) + 2 (2+3\sqrt{2})
I(2\bar\epsilon) - (2+3\sqrt{2}) I(4\bar\epsilon) +
I(8\bar\epsilon)}{3\sqrt{2}-5} \nonumber \\
R_4 (\bar\epsilon) & = & \frac{-32 I(\bar\epsilon) + 24(1+\sqrt{2})
I(2\bar\epsilon) - 6(2+3\sqrt{2}) I(4\bar\epsilon) +
3(2+\sqrt{2}) I(8\bar\epsilon) - I(16\bar\epsilon)}{3(3\sqrt{2}-5)}
\end{eqnarray}
and so forth. We used these polynomials in the
region of $\pm 1$ GeV around the $t\bar t$ threshold for the Richardson
extrapolation.

The on-shell Higgs wave function, vacuum expectation value and on-shell
top mass counterterms appear in the counterterms of the two-loop box
diagrams. They are given by
\begin{eqnarray}
\Delta_{\Box,CT} = \delta Z_H - 2\frac{\delta v}{v} - 2 \delta m_t
\frac{1}{F_\Box} \frac{\partial F_\Box}{\partial m_t} \nonumber \\
\Delta_{G,CT} = \delta Z_H - 2\frac{\delta v}{v} - 2 \delta m_t
\frac{1}{G_\Box} \frac{\partial G_\Box}{\partial m_t} \nonumber \\
\end{eqnarray}
for the two form factors $F_\Box$ and $G_\Box$, respectively. The final
results are summarized by the finite correction factors $\Delta_\Box,
\Delta_G$ of Eq.~(\ref{eq:corr}). 
It is known that in the HTL, $\Delta_\Box \to 4 x_t$ \cite{schlenk}, and we have checked numerically
that this limit is reproduced by our numerical calculation within
numerical errors by artificially increasing the size of the virtual top
mass to 3 TeV.

\subsection{Light-Quark-induced corrections}
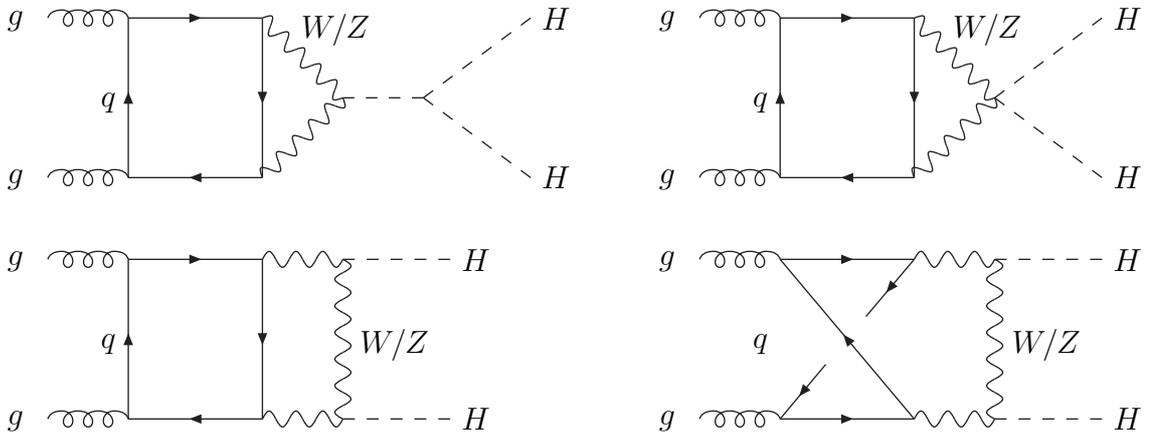
\begin{figure}[hbtp]
\begin{center}
\setlength{\unitlength}{1pt}
\begin{picture}(100,90)(120,0)
\Gluon(20,80)(50,80){3}{3}
\Gluon(20,20)(50,20){3}{3}
\ArrowLine(50,20)(50,80)
\ArrowLine(50,80)(100,80)
\ArrowLine(100,80)(100,20)
\ArrowLine(100,20)(50,20)
\Photon(100,80)(130,50){3}{5}
\Photon(100,20)(130,50){3}{5}
\DashLine(130,50)(160,50){5}
\DashLine(160,50)(200,80){5}
\DashLine(160,50)(200,20){5}
\put(205,76){$H$}
\put(205,16){$H$}
\put(5,78){$g$}
\put(5,18){$g$}
\put(40,47){$q$}
\put(115,73){$W/Z$}
\end{picture}
\begin{picture}(100,90)(-20,0)
\Gluon(20,80)(50,80){3}{3}
\Gluon(20,20)(50,20){3}{3}
\ArrowLine(50,20)(50,80)
\ArrowLine(50,80)(100,80)
\ArrowLine(100,80)(100,20)
\ArrowLine(100,20)(50,20)
\Photon(100,80)(130,50){3}{5}
\Photon(100,20)(130,50){3}{5}
\DashLine(130,50)(170,80){5}
\DashLine(130,50)(170,20){5}
\put(175,76){$H$}
\put(175,16){$H$}
\put(5,78){$g$}
\put(5,18){$g$}
\put(40,47){$q$}
\put(115,73){$W/Z$}
\end{picture} \\
\begin{picture}(100,90)(120,0)
\Gluon(20,80)(50,80){3}{3}
\Gluon(20,20)(50,20){3}{3}
\ArrowLine(50,20)(50,80)
\ArrowLine(50,80)(100,80)
\ArrowLine(100,80)(100,20)
\ArrowLine(100,20)(50,20)
\Photon(100,80)(130,80){3}{3}
\Photon(130,80)(130,20){3}{6}
\Photon(100,20)(130,20){3}{3}
\DashLine(130,80)(170,80){5}
\DashLine(130,20)(170,20){5}
\put(175,76){$H$}
\put(175,16){$H$}
\put(5,78){$g$}
\put(5,18){$g$}
\put(40,47){$q$}
\put(137,45){$W/Z$}
\end{picture}
\begin{picture}(100,90)(-20,0)
\Gluon(20,80)(50,80){3}{3}
\Gluon(20,20)(50,20){3}{3}
\ArrowLine(50,80)(100,80)
\ArrowLine(100,80)(82,58.5)
\ArrowLine(67,40.5)(50,20)
\ArrowLine(100,20)(50,80)
\ArrowLine(50,20)(100,20)
\Photon(100,80)(130,80){3}{3}
\Photon(130,80)(130,20){3}{6}
\Photon(100,20)(130,20){3}{3}
\DashLine(130,80)(170,80){5}
\DashLine(130,20)(170,20){5}
\put(175,76){$H$}
\put(175,16){$H$}
\put(5,78){$g$}
\put(5,18){$g$}
\put(40,47){$q$}
\put(137,45){$W/Z$}
\end{picture} \\[-0.3cm]
\setlength{\unitlength}{1pt}
\caption{\label{fg:lightquark} \it Sample two-loop box diagrams of the
light-quark-induced electroweak corrections to Higgs-boson pair
production. Shown are diagrams of the three classes -- triangle diagrams,
four-point diagrams and genuine planar/non-planar double boxes.}
\end{center}
\end{figure}

The second class of electroweak corrections we have calculated is the
light-quark-loop induced corrections. 
We take into account closed loops of all light-quark flavors up to the bottom quark in diagrams involving $Z$ bosons, and up to the charm quark in diagrams involving $W$ bosons, in order to avoid top-quark contributions. Also, we do \textit{not} work in the gaugeless limit in this case.
For single-Higgs production, these types of corrections
were the dominant part of the full NLO electroweak corrections
\cite{aglietti,passarino}. The corresponding Higgs-pair diagrams are split
into three different classes: the triangle diagrams with an $s$-channel
Higgs propagator, the triangle-like diagrams with a four-point $WWHH$
and $ZZHH$ interaction and the genuine double-box contributions, see
Fig.~\ref{fg:lightquark}.

\begin{figure}[hbt]
\begin{center}
\begin{picture}(150,245)(0,0)
\put(-150,-155.0){\includegraphics{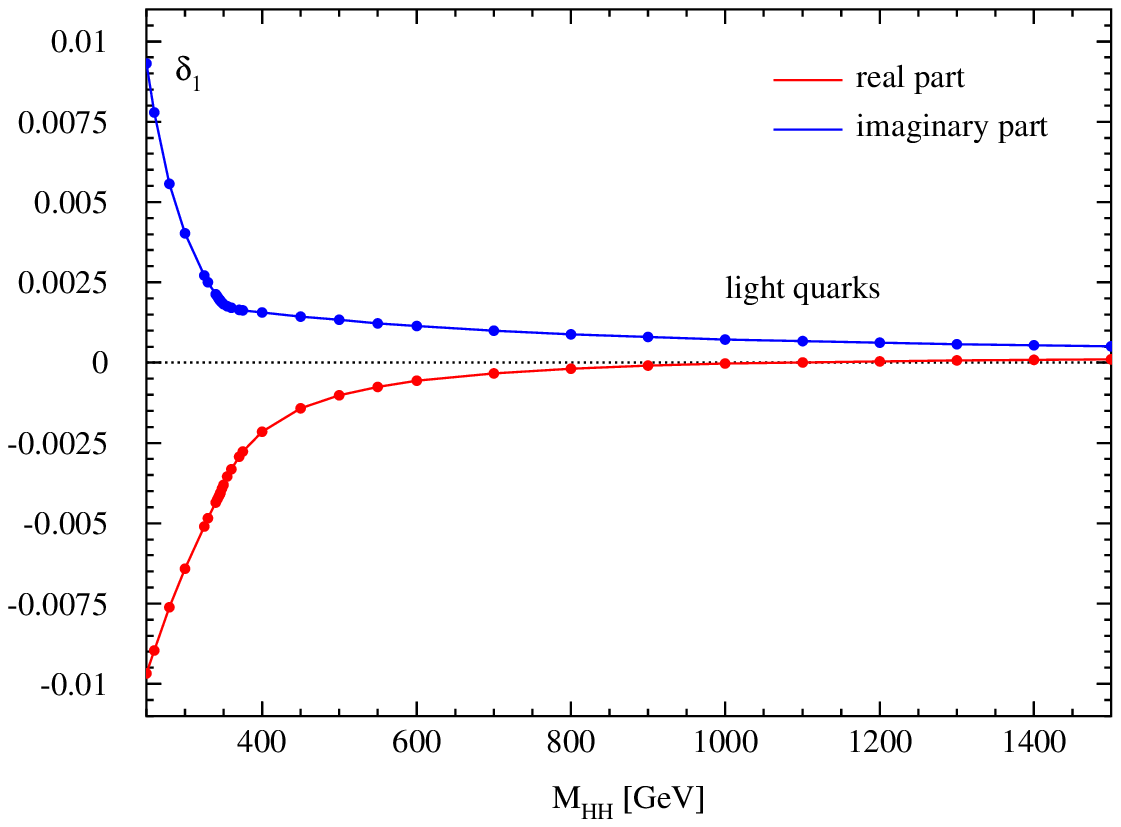}}
\end{picture}
\caption{\it The relative light-quark-induced electroweak correction
factor $\delta_1$ as a function of the invariant Higgs-pair mass
$M_{HH}$. The dots along the curves represent the numerical numbers we
have generated. Their error bars are negligible and thus not shown.}
\label{fg:d1_light}
\end{center}
\end{figure}

We have applied the same method as for the top-Yukawa induced diagrams
for these corrections, i.e.~suitable Feynman parametrizations of the
individual diagrams after projecting onto the two form factors, end-point
subtractions to separate the divergencies and integration-by-parts to
stabilize the virtual thresholds, where we introduced complex propagator
masses (including the $W$ and $Z$ boson masses, $M_{W/Z}^2 \to
M_{W/Z}^2 (1-i\bar\epsilon)$) as in Eq.~(\ref{eq:imaginary}). We have
explicitly checked that there are no infrared or collinear divergences
in the individual diagrams, so end-point subtractions alone are sufficient
to isolate the divergences. The UV poles cancel
between the planar and non-planar box diagrams, as well as for the
triangle and four-point diagrams individually. As a quantitative
cross-check, we implemented the corresponding results of the single Higgs
case \cite{aglietti} and compared them against our results for the
triangle diagrams and the four-point diagrams and found full agreement
with our numerical implementation. For the sake of accuracy and
stability, we computed the numerical results with these
implementations. For the genuine double-box diagrams, the dependence on
the imaginary parameter $\bar\epsilon$ is mild; therefore, we computed the results with the value $\bar\epsilon$ = 0.05, as this has been found to be in the plateau of the narrow-width limit $\bar\epsilon\to 0$ within integration errors.  The results of the triangle diagrams can be
accommodated by the same correction factor $\delta_1$ as in the
top-Yukawa induced part, see Eq.~(\ref{eq:triacorr}). The emerging
contribution $\delta_{1,lq}$ is shown in Fig.~\ref{fg:d1_light} and
amounts to less than a per-cent in total.

\section{Results} \label{sc:results}
Now, we are in the position to present the relative top-Yukawa and
light-quark induced electroweak corrections to Higgs-pair production via
gluon fusion.

\begin{figure}[hbtp]
\begin{center}
\begin{picture}(150,550)(0,0)
\put(-150,110.0){\includegraphics{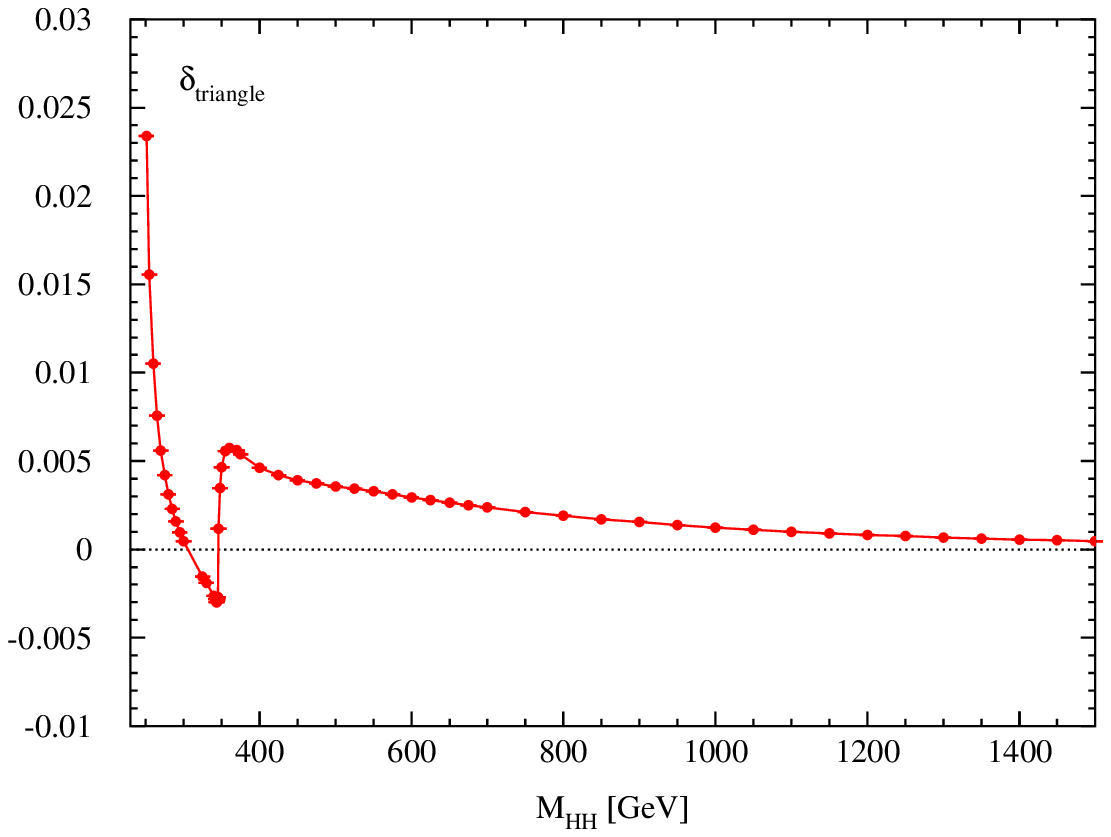}}
\put(-150,-155.0){\includegraphics{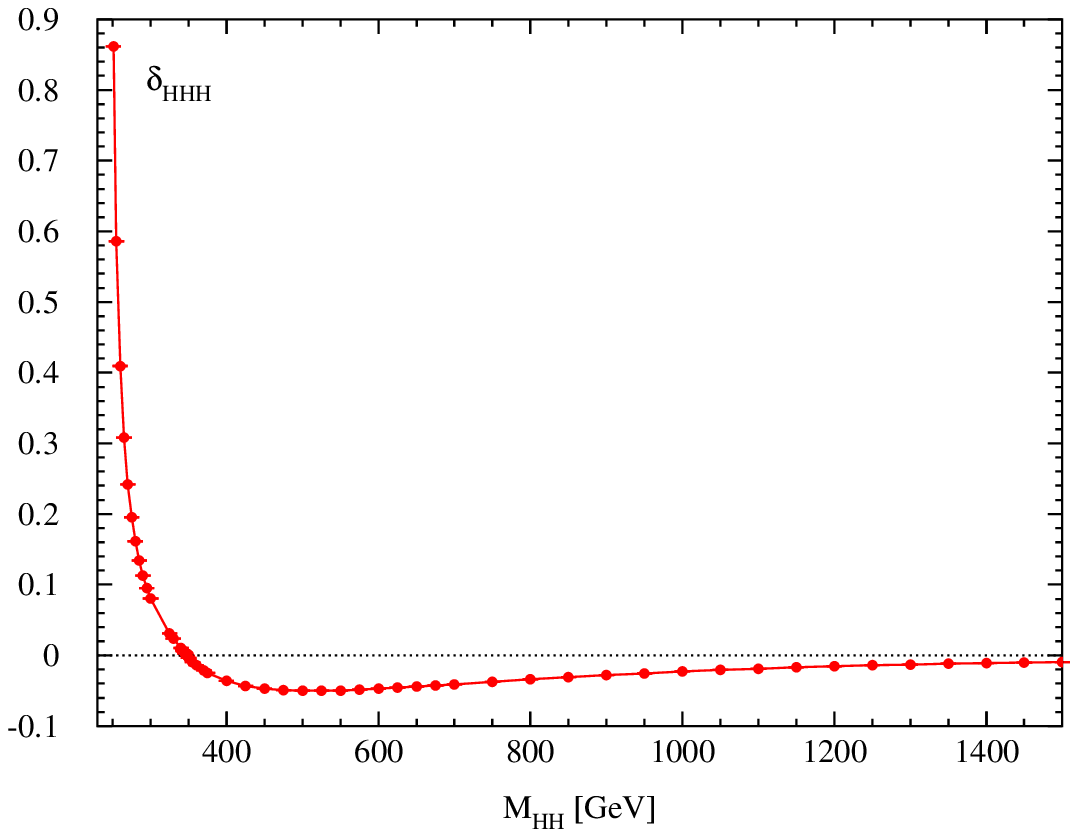}}
\end{picture}
\caption{\it The relative two-loop triangle (upper plot) and one-loop
times one-loop (lower plot) contributions to the
electroweak corrections of the Higgs-pair production cross section as a
function of the invariant Higgs-pair mass $M_{HH}$. The dots along the
curves represent the numerical numbers we have generated.}
\label{fg:delta_tria}
\end{center}
\end{figure}
\begin{figure}[hbtp]
\begin{center}
\begin{picture}(150,550)(0,0)
\put(-150,110.0){\includegraphics{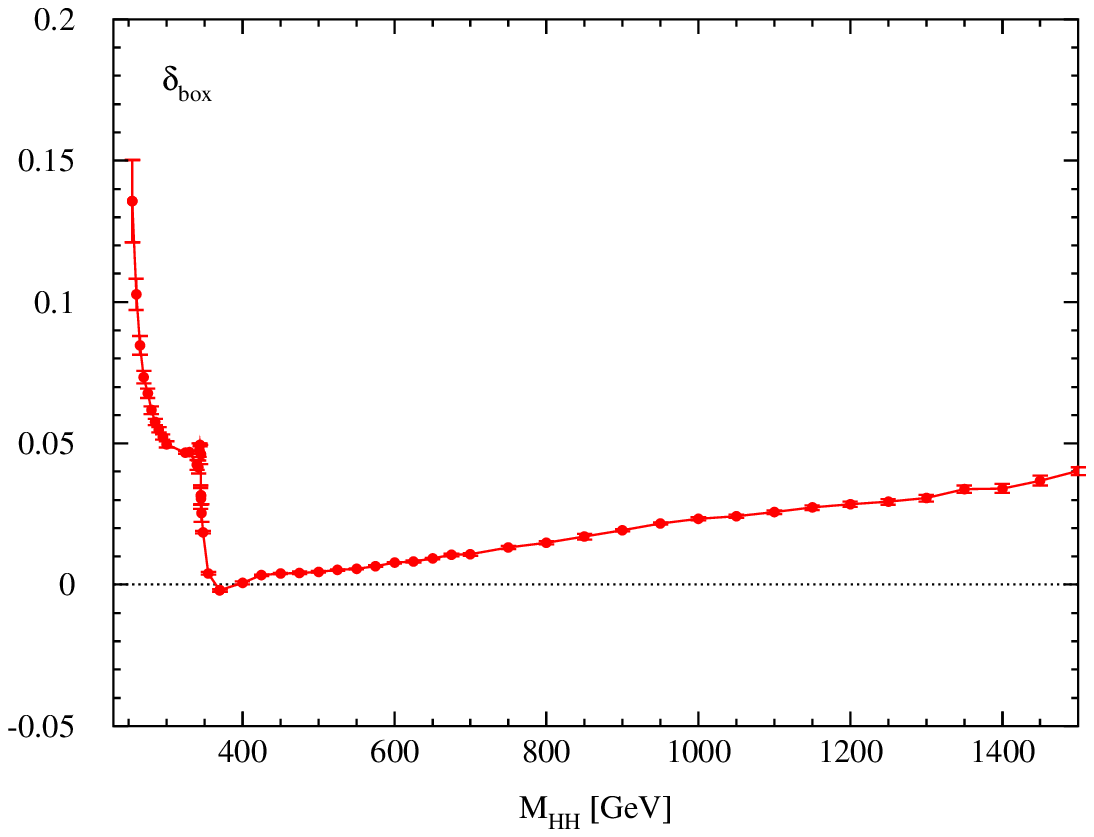}}
\put(-150,-155.0){\includegraphics{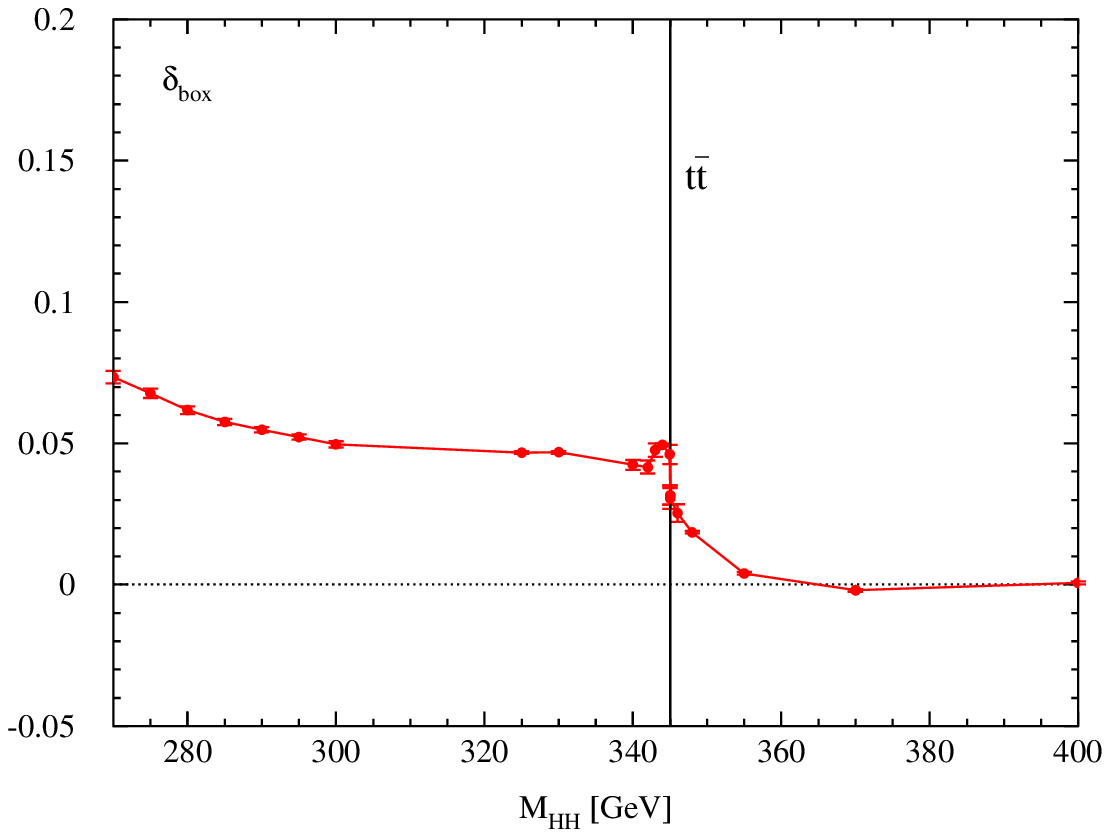}}
\end{picture}
\caption{\it The relative two-loop box contribution to the electroweak
corrections of the Higgs-pair production cross section as a function of
the invariant Higgs-pair mass $M_{HH}$. The lower plot shows the magnified region around the $t\bar t$ threshold. The dots along the curves
represent the numerical numbers we have generated and their error bars.}
\label{fg:delta_box}
\end{center}
\end{figure}
\begin{figure}[hbt]
\begin{center}
\begin{picture}(150,240)(0,0)
\put(-150,-155.0){\includegraphics{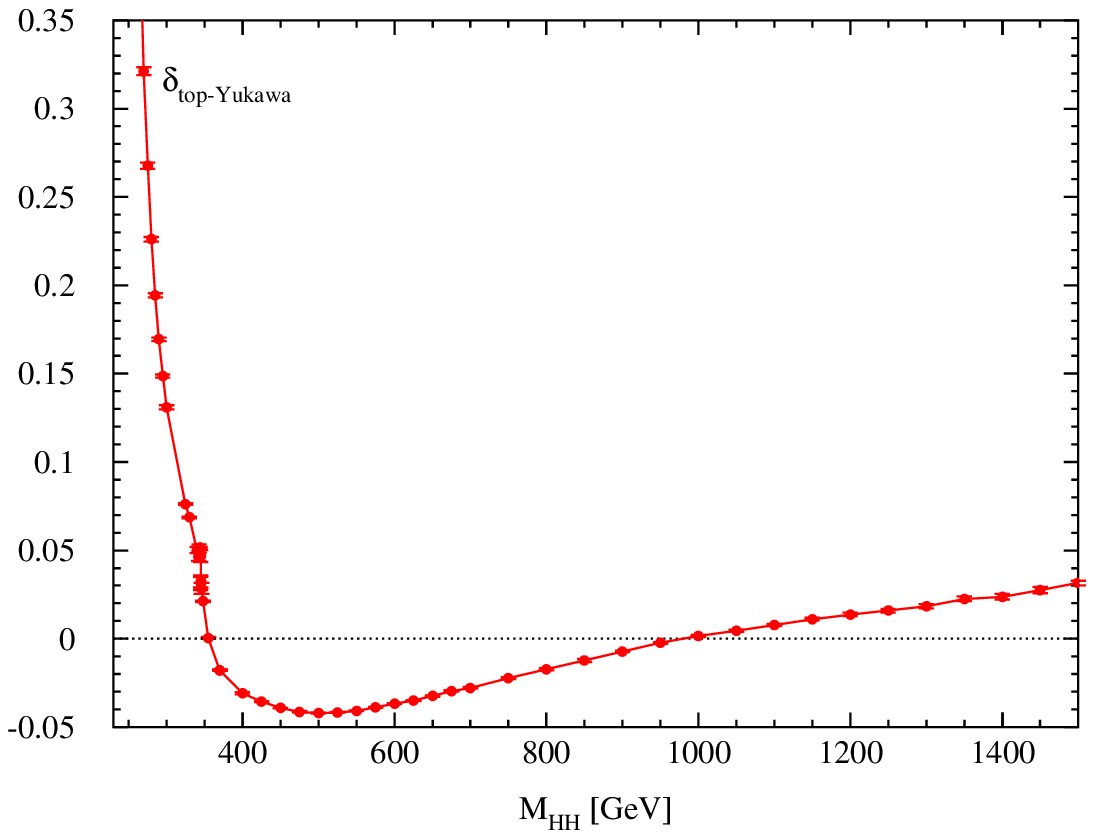}}
\end{picture}
\caption{\it The relative total two-loop top-Yukawa induced contribution
to the electroweak corrections of the Higgs-pair production cross
section as a function of the invariant Higgs-pair mass $M_{HH}$. The
dots along the curves represent the numerical numbers we have
generated and their error bars.}
\label{fg:delta_t-yuk}
\end{center}
\end{figure}

Before we move to the total electroweak corrections $\delta$ of
Eq.~(\ref{eq:delta}), we discuss the individual contributions of the
top-Yukawa and light-quark induced components,
\begin{eqnarray}
\delta & = & \delta_{top-Yukawa} + \delta_{light-quarks} \nonumber \\
\delta_{top-Yukawa} & = & \delta_{triangle} + \delta_{HHH} + \delta_{box}
\end{eqnarray}
In Fig.~\ref{fg:delta_tria}, the individual contributions of the two-loop
triangle and one-loop times one-loop diagrams are shown. The size of the
triangle-diagram part ranks at the per-cent level in
most of the invariant Higgs-pair mass range, apart from the region close
to the production threshold $M_{HH}\gsim 2M_H$. In this region, the corrections
are larger, since they are lifting the strong cancelation of
the triangle and box diagrams at LO. On the other hand, the contribution $\delta_{HHH}$
emerging from the one-loop times one-loop diagrams, $\Delta_{HHH}$ of
Eq.~(\ref{eq:triacorr}), is significantly larger, at the --5\% level for
larger invariant Higgs-pair masses $M_{HH}$ as displayed in the lower part of Fig.~\ref{fg:delta_tria}. These corrections provide a sizable
correction close to the production threshold, which is in line with the
results of Ref.~\cite{schlenk}. The contribution of the two-loop box
diagrams, as depicted in Fig.~\ref{fg:delta_box}, is at the 5--10\%
level. The electroweak corrections are dominated by the contributions of form factor $F_1$ of Eq.~(\ref{eq:ff}), while form factor $F_2$ is only relevant for larger values of $M_{HH}$.
 To properly describe the $t\bar{t}$ threshold behavior, a deeper analysis on the convergences of the Richardson polynomials has been performed. In contrast to the other regions, the amplitudes are strongly dependent on $\bar{\epsilon}$. Thus, to ensure the narrow-width limit, the Richardson polynomial needs a minimal regulator $\bar{\epsilon} \sim 10^{-2}$.
 As the result, the virtual $t\bar t$ threshold appears as a shoulder in the invariant mass distribution with a maximum-minimum behaviour around the $t\bar t$ threshold. This is shown in detail in the lower plot of Fig.~\ref{fg:delta_box} and is in line with the ${\cal P}$-wave property of the threshold at LO that prevents a kink-like structure at NLO.
Summing up all ingredients for the total top-Yukawa induced corrections,
we arrive at the total relative corrections shown in
Fig.~\ref{fg:delta_t-yuk}. The total corrections amount to about $\pm
5\%$ in the large-$M_{HH}$ regime, while they are larger close to the
production threshold.

\begin{figure}[hbt]
\begin{center}
\begin{picture}(150,245)(0,0)
\put(-150,-155.0){\includegraphics{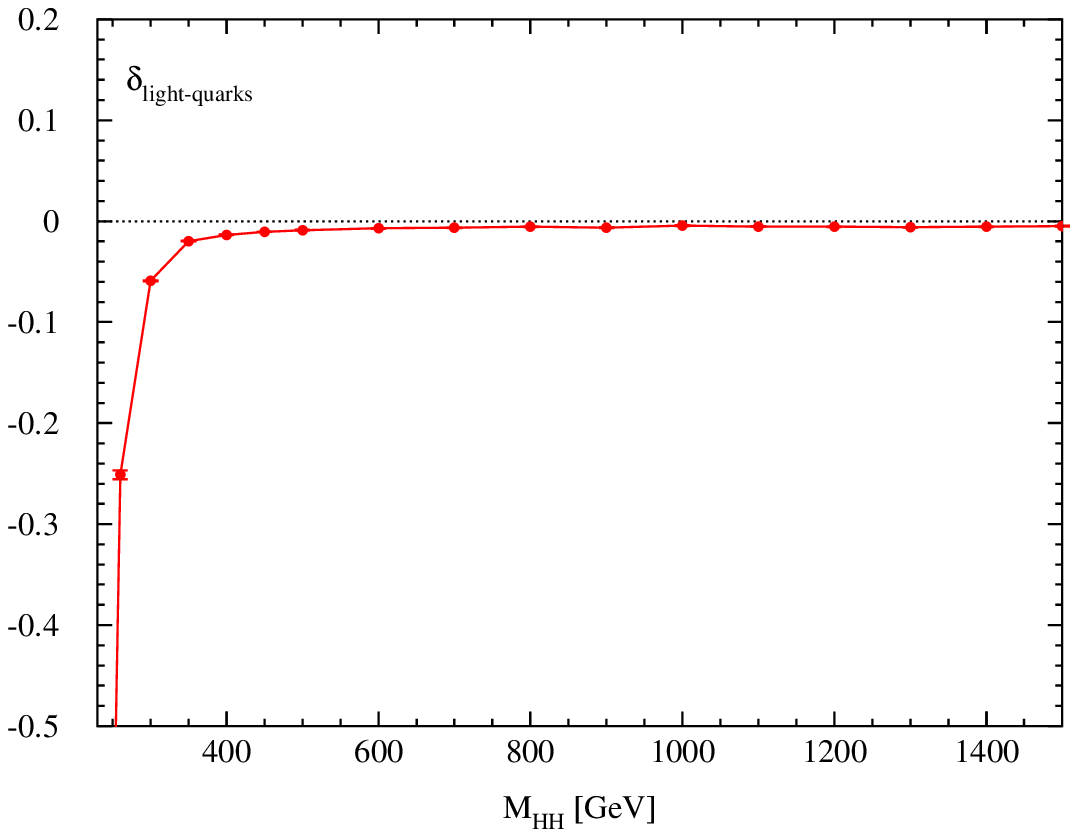}}
\end{picture}
\caption{\it The relative light-quark-induced contribution to the
electroweak corrections of the Higgs-pair production cross section as a
function of the invariant Higgs-pair mass $M_{HH}$. The dots along the
curves represent the numerical numbers we have generated and their error bars.}
\label{fg:delta_lq}
\end{center}
\end{figure}

As an additional contribution to the electroweak corrections, the total result of the light-quark induced electroweak corrections, denoted by $\delta_{\text{light-quarks}}$, is shown in Fig.~\ref{fg:delta_lq}. We implemented a Richardson extrapolation for two values of the imaginary regulator, $\bar\epsilon = 0.1$ and $0.05$, which proved sufficient to reach the narrow-width limit across the full range of $M_{HH}$.
These corrections are dominated by the form factor $F_1$ of Eq.~(\ref{eq:ff}), while the form factor $F_2$ becomes relevant only for larger values of the invariant Higgs-pair mass $M_{HH}$. The light-quark
contributions are small for large invariant Higgs-pair masses $M_{HH}$
and are relevant only close to the production threshold, where they are negative.

Using a suitable interpolation of our results implemented in the code {\tt Hpair} \cite{hpair} we arrive at the individual corrections to
the total hadronic cross section that amount to $-1.9\%$ for the top-Yukawa induced part and $-1.5\%$ for the light-quark induced corrections. In total, our corrections as part of the full electroweak~corrections reduce the total cross section by $3.4\%$.

\section{Conclusions} \label{sc:conclusions}
In this work, we calculated the top-Yukawa and light-quark induced electroweak corrections to Higgs-pair production via gluon fusion, $gg\to HH$. We applied the same numerical methods as in the case of the QCD corrections at NLO \cite{gghhnlo1,gghhqcd2}, i.e.~projection on the two physical form factors of the spin-0 and 2 contributions, Feynman parametrization of all two-loop diagrams individually, end-point subtractions to separate the ultraviolet divergences and integration-by-parts to stabilize the numerical integrals across the virtual thresholds. Above the virtual thresholds, we introduced complex propagator masses with a small imaginary part and used Richardson extrapolations to arrive at the narrow-width limit. The total top-Yukawa induced corrections amount to the level of 5--10\% for moderate and large values of the invariant Higgs-pair mass $M_{HH}$, while they are larger close to the production threshold $M_{HH}\gsim 2M_H$ due to the strong destructive interference of the LO matrix element. The top-Yukawa induced corrections develop a shoulder at the virtual $t\bar t$ threshold that is in line with the ${\cal P}$-wave nature of this threshold at LO. For larger values of $M_{HH}$, we observe a finite slope in the relative corrections that modify the differential cross section by a few per-cent up to the 5\%-level. On the other hand, the light-quark induced corrections are tiny in the large $M_{HH}$ range and only contribute significantly close to the production threshold, again, due to strong suppression of the LO matrix element. The individual corrections to the total hadronic cross section amount to $-1.9\%$ for the top-Yukawa induced part and $-1.5\%$ for the light-quark induced corrections. This work presents the first steps towards the calculation of the full electroweak corrections. \\

\noindent
{\bf Acknowledgments} \\
The authors are indebted to M.~Bonetti, C.~Borschensky, G.~Heinrich, S.~Jones, M.~Kerner, P.~Rendler, and A.~Vestner for useful discussions. We would like to thank M.~Bonetti and P.~Rendler for comparing explicit numbers of the light-quark induced corrections. The work of A.B. and F.C. is supported by the Generalitat Valenciana, the Spanish government, and ERDF funds from the European Commission “NextGenerationEU/PRTR” (CNS2022-136165, PID2023-151418NB-I00, MCIN/AEI/10.13039/501100011033/). The research of S.C.~and M.M.~is supported by the Deutsche Forschungsgemeinschaft (DFG, German Research Foundation) under grant 396021762--TRR 257. The work of J.C.~is supported by the Swiss National Science Foundation (SNSF). J.R.
acknowledges support from INFN. We acknowledge support by the state of Baden-W\"urttemberg through bwHPC and the German Research Foundation (DFG) through grant no INST 39/963-1 FUGG (bwForCluster NEMO).



\begin{thebibliography}{99}

\bibitem{discovery}
  G.~Aad {\it et al.} [ATLAS Collaboration],
  Phys.\ Lett.\ {\bf B716} (2012) 1;
  S.~Chatrchyan {\it et al.} [CMS Collaboration],
  Phys.\ Lett.\ {\bf B716} (2012) 30.

\bibitem{couplings}
  G.~Aad {\it et al.} [ATLAS and CMS Collaborations],
  JHEP {\bf 1608} (2016) 045;
 G.~Aad {\it et al.} [ATLAS Collaboration],
ATLAS-CONF-2019-005;
A.M.~Sirunyan \textit{et al.} [CMS Collaboration],
JHEP \textbf{01} (2021) 148.

\bibitem{h2zgamma}
G.~Aad \textit{et al.} [ATLAS],
Phys. Lett. B \textbf{809} (2020), 135754
and
  arXiv:2507.12598 [hep-ex];
A.~Tumasyan \textit{et al.} [CMS],
JHEP \textbf{05} (2023), 233.

\bibitem{hhpro}
A.~Djouadi, W.~Kilian, M.~M\"uhlleitner and P.M.~Zerwas,
Eur.~Phys.~J.~\textbf{C10} (1999), 45.

\bibitem{higgs}
  P.~W.~Higgs,
  Phys.\ Lett.\  {\bf 12} (1964) 132,
  Phys.\ Rev.\ Lett.\  {\bf 13} (1964) 508
and
  Phys.\ Rev.\  {\bf 145} (1966) 1156;
  F.~Englert and R.~Brout,
  Phys.\ Rev.\ Lett.\  {\bf 13} (1964) 321;
  G.~S.~Guralnik, C.~R.~Hagen and T.~W.~Kibble,
  Phys.\ Rev.\ Lett.\  {\bf 13} (1964) 585;
  T.~W.~B.~Kibble,
  Phys.\ Rev.\  {\bf 155} (1967) 1554.

\bibitem{unitarity}
  C.~H.~Llewellyn Smith,
  Phys.\ Lett.\  {\bf 46B} (1973) 233;
  J.~M.~Cornwall, D.~N.~Levin and G.~Tiktopoulos,
  Phys.\ Rev.\ D {\bf 10} (1974) 1145
   Erratum: [Phys.\ Rev.\ D {\bf 11} (1975) 972];
  B.~W.~Lee, C.~Quigg and H.~B.~Thacker,
  Phys.\ Rev.\ Lett.\  {\bf 38} (1977) 883
and
  Phys.\ Rev.\ D {\bf 16} (1977) 1519.

\bibitem{smren}
  G.~'t Hooft,
  Nucl.\ Phys.\ B {\bf 35} (1971) 167;
  G.~'t Hooft and M.~J.~G.~Veltman,
  Nucl.\ Phys.\ B {\bf 44} (1972) 189.

\bibitem{review}
  M.~Spira,
  Fortsch.\ Phys.\  {\bf 46} (1998) 203
and
  Prog.\ Part.\ Nucl.\ Phys.\  {\bf 95} (2017) 98;
A.~Djouadi,
Phys. Rept. \textbf{457} (2008), 1-216.

\bibitem{hhrev}
  J.~Baglio, A.~Djouadi, R.~Gr\"ober, M.~M.~M\"uhlleitner, J.~Quevillon and
  M.~Spira,
  JHEP {\bf 1304} (2013) 151;
  B.~Di Micco, M.~Gouzevitch, J.~Mazzitelli, C.~Vernieri, J.~Alison,
  K.~Androsov, J.~Baglio, E.~Bagnaschi, S.~Banerjee and P.~Basler,
  \textit{et al.},
  Rev.~Phys.~\textbf{5} (2020) 100045.

\bibitem{gghhlo}
  E.~W.~N.~Glover and J.~J.~van der Bij,
  Nucl.\ Phys.\ {\bf B309} (1988) 282;
  T.~Plehn, M.~Spira and P.~M.~Zerwas,
  Nucl.\ Phys.\ {\bf B479} (1996) 46,
   Erratum: [Nucl.\ Phys.\ {\bf B531} (1998) 655].

\bibitem{nlohtl}
S.~Dawson, S.~Dittmaier and M.~Spira,
Phys.~Rev.~\textbf{D58} (1998) 115012.

\bibitem{nnlohtl}
  D.~de Florian and J.~Mazzitelli,
  Phys.\ Lett.\ B {\bf 724} (2013) 306
and
  Phys.\ Rev.\ Lett.\  {\bf 111} (2013) 201801;
  J.~Grigo, K.~Melnikov and M.~Steinhauser,
  Nucl.\ Phys.\ B {\bf 888} (2014) 17.

\bibitem{n3lohtl}
  L.~B.~Chen, H.~T.~Li, H.~S.~Shao and J.~Wang,
  Phys.~Lett.~{\bf B803} (2020) 135292
and
  JHEP {\bf 2003} (2020) 072.

\bibitem{gghhnlo}
  S.~Borowka, N.~Greiner, G.~Heinrich, S.~P.~Jones, M.~Kerner,
  J.~Schlenk, U.~Schubert and T.~Zirke,
  Phys.\ Rev.\ Lett.\  {\bf 117} (2016) no.1,  012001
   Erratum: [Phys.\ Rev.\ Lett.\  {\bf 117} (2016) no.7,  079901];
  S.~Borowka, N.~Greiner, G.~Heinrich, S.~P.~Jones, M.~Kerner,
  J.~Schlenk and T.~Zirke,
  JHEP {\bf 1610} (2016) 107.

\bibitem{gghhnlo1}
J.~Baglio, F.~Campanario, S.~Glaus, M.~M\"uhlleitner, M.~Spira and
J.~Streicher,
Eur.~Phys.~J.~\textbf{C79} (2019) no.6, 459;
J.~Baglio, F.~Campanario, S.~Glaus, M.~M\"uhlleitner, J.~Ronca and
M.~Spira,
Phys.~Rev.~\textbf{D103} (2021) no.5, 056002.

\bibitem{gghhqcd2}
J.~Baglio, F.~Campanario, S.~Glaus, M.~M\"uhlleitner, J.~Ronca, M.~Spira
and J.~Streicher,
JHEP \textbf{04} (2020), 181.

\bibitem{gghhexp}
J.~Grigo, K.~Melnikov and M.~Steinhauser,
Nucl. Phys. B \textbf{888} (2014), 17-29;
J.~Grigo, J.~Hoff and M.~Steinhauser,
Nucl. Phys. B \textbf{900} (2015), 412-430;
R.~Gr{\"o}ber, A.~Maier and T.~Rauh,
JHEP \textbf{03} (2018), 020;
R.~Bonciani, G.~Degrassi, P.~P.~Giardino and R.~Gr{\"o}ber,
Phys. Rev. Lett. \textbf{121} (2018) no.16, 162003;
J.~Davies, G.~Mishima, M.~Steinhauser and D.~Wellmann,
JHEP \textbf{01} (2019), 176;
J.~Davies, G.~Heinrich, S.~P.~Jones, M.~Kerner, G.~Mishima, M.~Steinhauser
and D.~Wellmann,
JHEP \textbf{11} (2019), 024;
L.~Bellafronte, G.~Degrassi, P.~P.~Giardino, R.~Gr{\"o}ber and M.~Vitti,
JHEP \textbf{07} (2022), 069.

\bibitem{gghhexp1}
E.~Bagnaschi, G.~Degrassi and R.~Gr{\"o}ber,
Eur. Phys. J. C \textbf{83} (2023) no.11, 1054.

\bibitem{nnloftapprox}
D.~de Florian and J.~Mazzitelli, Phys.~Lett.~{\bf B724} (2013) 306 and
Phys.~Rev.~Lett.~{\bf 111} (2013) 201801;
J.~Grigo et al., Nucl.~Phys.~{\bf B888} (2014) 17;
  M.~Grazzini, G.~Heinrich, S.~Jones, S.~Kallweit, M.~Kerner,
  J.~M.~Lindert and J.~Mazzitelli,
  JHEP {\bf 1805} (2018) 059.

\bibitem{gghhresum}
D.~Y.~Shao, C.~S.~Li, H.~T.~Li and J.~Wang,
JHEP \textbf{07} (2013), 169;
D.~de Florian and J.~Mazzitelli,
JHEP \textbf{09} (2015), 053
and
JHEP \textbf{08} (2018), 156;
A.~H.~Ajjath and H.~S.~Shao,
JHEP \textbf{02} (2023), 067.

\bibitem{gghhpowheg}
G.~Heinrich, S.~P.~Jones, M.~Kerner, G.~Luisoni and E.~Vryonidou,
JHEP \textbf{08} (2017), 088;
S.~Jones and S.~Kuttimalai,
JHEP \textbf{02} (2018), 176.

\bibitem{gghhgeneva}
S.~Alioli, G.~Marinelli and D.~Napoletano,
arXiv:2507.08558 [hep-ph].

\bibitem{gghhnnlo}
J.~Davies, K.~Sch{\"o}nwald and M.~Steinhauser,
Phys. Lett. B \textbf{845} (2023), 138146
and
[arXiv:2503.17449 [hep-ph]];
J.~Davies, K.~Sch{\"o}nwald, M.~Steinhauser and M.~Vitti,
JHEP \textbf{08} (2024), 096.

\bibitem{gghhscet}
S.~Jaskiewicz, S.~Jones, R.~Szafron and Y.~Ulrich,
[arXiv:2501.00587 [hep-ph]].

\bibitem{gaugeless}
J.~R.~Espinosa and R.~J.~Zhang,
Nucl. Phys. B \textbf{586} (2000), 3-38;
A.~Brignole, G.~Degrassi, P.~Slavich and F.~Zwirner,
Nucl. Phys. B \textbf{631} (2002), 195-218.

\bibitem{schlenk}
M.~M{\"u}hlleitner, J.~Schlenk and M.~Spira,
JHEP \textbf{10} (2022), 185.

\bibitem{gghhelw}
H.~Y.~Bi, L.~H.~Huang, R.~J.~Huang, Y.~Q.~Ma and H.~M.~Yu,
Phys. Rev. Lett. \textbf{132} (2024) no.23, 231802.

\bibitem{gghhhiggs}
G.~Heinrich, S.~Jones, M.~Kerner, T.~Stone and A.~Vestner,
JHEP \textbf{11} (2024), 040.

\bibitem{gghhelwexp}
J.~Davies, G.~Mishima, K.~Sch{\"o}nwald, M.~Steinhauser and H.~Zhang,
JHEP \textbf{08} (2022), 259,
JHEP \textbf{10} (2023), 033
and
JHEP \textbf{04} (2025), 193.

\bibitem{bonetti}
M.~Bonetti, P.~Rendler and W.~J.~Torres Bobadilla,
JHEP \textbf{07} (2025), 024.

\bibitem{hooftvelt}
G.~'t Hooft and M.~J.~G.~Veltman,
Nucl. Phys. B \textbf{153} (1979), 365-401.

\bibitem{ggasqcd}
E.~Bagnaschi, L.~Fritz, S.~Liebler, M.~M{\"u}hlleitner, T.~T.~D.~Nguyen and
M.~Spira,
JHEP \textbf{03} (2023), 124;
L.~Fritz,
PhD thesis, University of Zurich, 2023.

\bibitem{aglietti}
  U.~Aglietti, R.~Bonciani, G.~Degrassi and A.~Vicini,
  Phys.\ Lett.\ B {\bf 595} (2004) 432
and
  hep-ph/0610033;
  G.~Degrassi and F.~Maltoni,
  Phys.\ Lett.\ B {\bf 600} (2004) 255.
R.~Bonciani, G.~Degrassi and A.~Vicini,
Comput. Phys. Commun. \textbf{182} (2011), 1253-1264.

\bibitem{passarino}
  S.~Actis, G.~Passarino, C.~Sturm and S.~Uccirati,
  Phys.\ Lett.\ B {\bf 670} (2008) 12;
  S.~Actis, G.~Passarino, C.~Sturm and S.~Uccirati,
  Nucl.\ Phys.\ B {\bf 811} (2009) 182.

\bibitem{hpair}
{\tt https://gitea.psi.ch/ltpth/HPAIR}

\end{thebibliography}
\end{document}